\documentclass[fleqn,usenatbib]{mnras}
\usepackage{newtxtext,newtxmath}

\usepackage[T1]{fontenc}

\usepackage{natbib}
\usepackage{graphicx}	
\usepackage{amsmath}
\usepackage{wasysym}           
\usepackage{mathrsfs}
\usepackage{anyfontsize}
\usepackage{mathtools}  
\usepackage{diffcoeff} 
\usepackage{color}
\usepackage{lipsum}
\usepackage{diagbox}
\usepackage{multicol}
\usepackage{caption}
\usepackage{multirow}
\usepackage{subcaption}
\usepackage[toc,page]{appendix}

\title[Dynamical stability of giant planets]{Dynamical stability of giant planets: the critical adiabatic index in the presence of a solid core}

\author[Kundu, Coughlin, Youdin, \& Armitage]{
Suman Kumar~Kundu$^{1}$\thanks{E-mail: skundu@syr.edu},
Eric R.~Coughlin$^{1}$\thanks{E-mail: ecoughli@syr.edu}, Andrew N.~Youdin$^{2}$\thanks{E-mail:youdin@arizona.edu}, Philip J.~Armitage${^{3,4}}$\thanks{E-mail:philip.armitage@stonybrook.edu}
\\
$^{1}$Department of Physics, Syracuse University, Syracuse, NY 13244, USA
\\
$^{2}$Steward Observatory, University of Arizona, 933 N. Cherry Avenue, Tucson, AZ 85721, USA
\\
$^{3}$Department of Physics and Astronomy, Stony Brook University, Stony Brook, NY 11794, USA 
\\
$^{4}$Center for Computational Astrophysics, Flatiron Institute, New York, NY 10010, USA
}

\date{Accepted XXX. Received YYY; in original form ZZZ}

\pubyear{2021}

\begin{document}
\label{firstpage}
\pagerange{\pageref{firstpage}--\pageref{lastpage}}
\maketitle
\begin{abstract}
The dissociation and ionization of hydrogen, during the formation of giant planets via core accretion, reduces the effective adiabatic index $\gamma$ of the gas and could trigger dynamical instability. We generalize the analysis of Chandrasekhar, who determined that the threshold for instability of a self-gravitating hydrostatic body lies at $\gamma=4/3$, to account for the presence of a planetary core, which we model as an incompressible fluid. We show that the dominant effect of the core is to stabilize the envelope to radial perturbations, in some cases completely (i.e. for all $\gamma > 1$). When instability is possible, unstable planetary configurations occupy a strip of $\gamma$ values whose upper boundary falls below $\gamma=4/3$. Fiducial evolutionary tracks of giant planets forming through core accretion appear unlikely to cross the dynamical instability strip that we define.
\end{abstract}

\begin{keywords}
Hydrodynamics --- instabilities --- methods: analytical
 --- planets and satellites: dynamical evolution and stability
\end{keywords}

\section{Introduction}
\label{sec:introduction}
One of the ways by which giant planets are thought to form is through ``core accretion'' (CA\footnote{The other paradigm for giant-planet formation is through instabilities in the protoplanetary disk -- the ``disc instability'' model (e.g., \citealt{cameron78,adams89, Boss97}). We will cast our discussion within the framework of the CA model, but many of our conclusions would also apply to the disk instability model. See \citet{2011ASPC..447...47K, nixon18} for discussions of these models.}; e.g., \citealt{safronov72, perri74, harris78, hiroshi80,STEVENSON1982755,Boden86}), a sequential process whereby the planet is assembled from a solid core that builds a surrounding, gaseous envelope through accretion from a protoplanetary disc.
Quantitative studies of the CA model show that it proceeds via two or three evolutionary phases \citep{pollack96}.
In the first phase planetesimals form \citep{Youdin_2005, doi:10.1146/annurev-earth-040809-152513, 2014prpl.conf..547J, 2019A&A...630A...2H} and then grow into larger rocky cores by the accretion of pebbles and/or other planetesimals and growing cores \citep{wetherill93, Goldreich_2004, 2007A&A...466..413O,2013pss3.book....1Y, 2019A&A...624A.114L}.
In the second phase the core accretes a low mass (relative to the core) gas envelope from the circumstellar disc by radiative cooling on the Kelvin-Helmholtz (KH) timescale, which causes the envelope to contract and permits further gaseous accretion \citep{2014ApJ...786...21P,Lee_2015}.
Initially, the KH timescale increases as the envelope grows in mass, and thus the evolution is slow. The recycling of gas between the disc and the bound envelope, a three-dimensional effect that is not captured in classical models, may modify the cooling timescale at this stage \citep{Machida_2008,Tanigawa_2012, 2017A&A...606A.146L}. If and when the atmospheric mass becomes comparable to that of the core, the self-gravity of the gas envelope becomes significant, and the KH timescale decreases. This transition leads to the third phase of rapid ``runaway growth" of the gas envelope, which (in classical one-dimensional models) begins as a more rapid, but still hydrostatic, KH contraction (e.g., \citealt{Boden86, 2000ApJ...537.1013I}.
The end of runaway growth again involves hydrodynamics, as the growing planet carves a gap in the disk, which lowers gas accretion and eventually sets the final planet mass \citep{Lubow_1999, 2009Icar..199..338L, ginz19}. 

Qualitative and potentially observable changes to the giant planet formation process would occur if, at any point, the planet became dynamically unstable. This is in principle possible if the effective adiabatic index of the gaseous envelope, which we denote as $\gamma$, drops below a critical value. \citet{chandra33} demonstrated that for a gaseous, self-gravitating spherical body, dynamical instability to radial perturbations occurs if $\gamma < 4/3$. This result has well-studied implications for the stability of massive stars, which are dominated by radiation pressure and hence have $\gamma \approx 4/3$, but it can also come into play for cooler, gas pressure dominated bodies. As a gas envelope contracts and its temperature rises, molecular hydrogen first dissociates, and is then ionized. These two transitions can reduce the adiabatic index from 7/5 (appropriate to a diatomic molecular gas) to values near unity \citep{saumon95}. In the star formation context is has long been suggested that this drop in adiabatic index as gas heats up leads to dynamical instability and collapse \citep{larson69}, and it is natural to ask whether the same phenomenon can occur in planet formation. Early work on core accretion noted that the existence of low $\gamma$ regions could bring proto-planets close to dynamical instability \citep{perri74}, which could lead to pulsations \citep{1991Icar...91...53W}. More recently, \citet{ginz19} speculated that the presence of $\gamma < 4/3$ regions in enough of the planet might accelerate planet formation. We emphasize that this potential {\em dynamical} instability is distinct from the established effect that dissociation has in reducing gas accretion rates during the slow, hydrostatic phase \citep{2015ApJ...800...82P}.

The key difference between the structure of a star and that of a forming giant planet is that the latter possesses a rocky or icy core that is much less compressible than the gaseous envelope. It is thus inaccurate to treat the entire planet as a single adiabatic gas, and Chandrasekhar's simple $\gamma < 4/3$ criterion for instability need not apply. A more realistic -- though still highly idealized -- model is to treat the planet as having a gaseous envelope of fixed adiabatic index, that is truncated radially at a finite inner core radius. Our goal in this paper is to calculate a modified Chandrasekhar-like stability criterion for this model system, which will depend on the size and mass of the rocky core. We consider the outer radius of the planet embedded in the disc to be less than the smaller of the Hill radius or the Bondi radius, and ignore any non-radial perturbation from the stellar tidal force.

In Section \ref{sec:equations} we describe the model and write down the fluid equations that govern the evolution of the planetary envelope in the presence of the massive core, and in Section \ref{sec:static} we present hydrostatic solutions to these equations in the limit that the envelope can be modeled as polytropic. In Section \ref{sec:perturbations} we derive the  equations that describe the response of the envelope to radial perturbations and we delimit the region of instability as a function of the core properties.  
We discuss the implications of our findings and directions for future work in Section \ref{sec:summary}.

\section{Atmospheric model} 
\label{sec:model}

\begin{figure} 
\centering
    \includegraphics[width=0.4\textwidth]{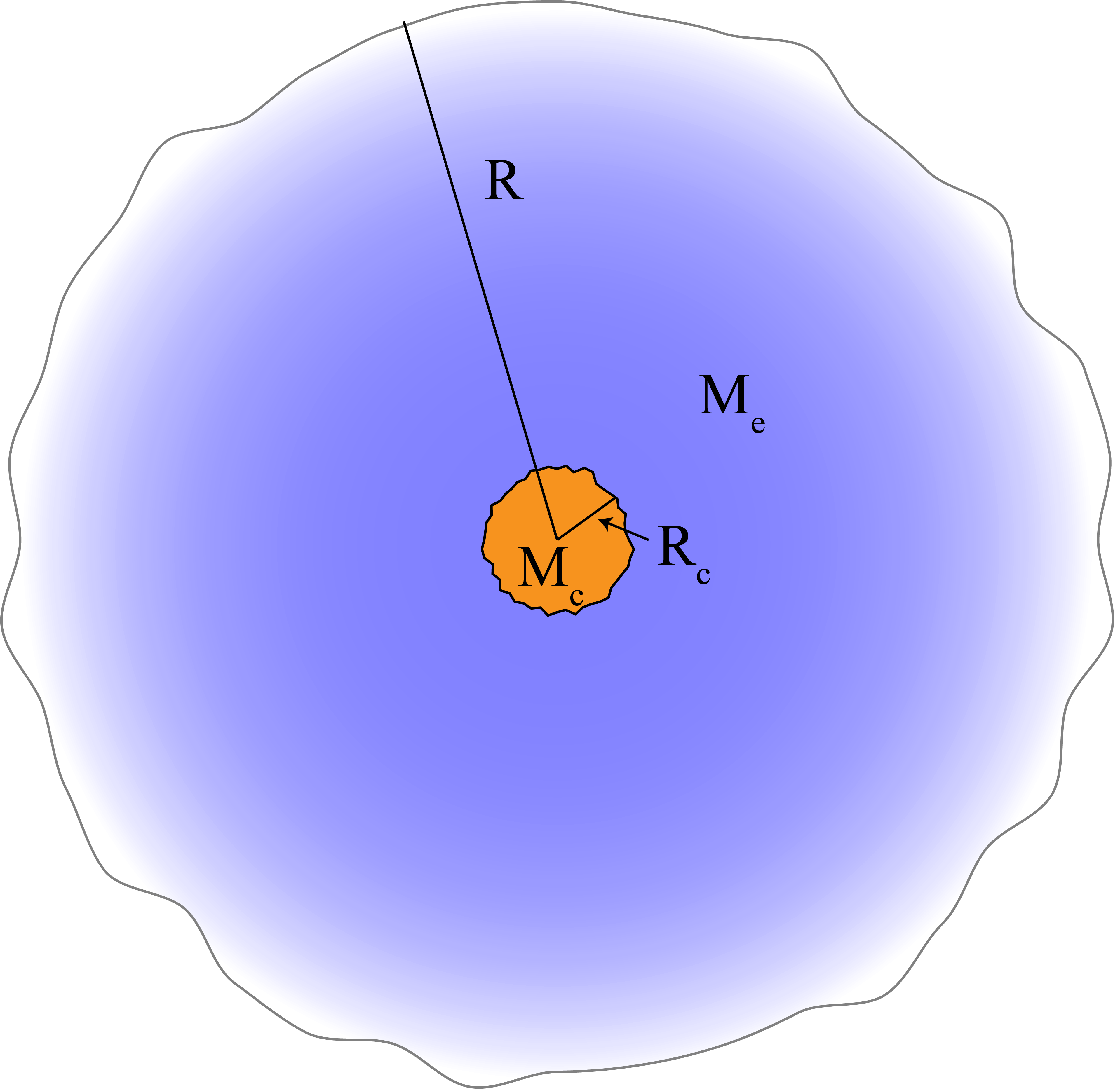}\par
    \caption{A qualitative illustration of our model of a giant planet, which is a spherical object that consists of a core of radius $R_{\rm c}$ and mass $M_{\rm c}$ and a compressible envelope that extends from $R_{\rm c}$ to $R$. The envelope has a mass of $M_{\rm e}$, such that the total mass of the planet is $M_{\rm c}+M_{\rm e}$ and the relative mass of the core is $\mu = M_{\rm c}/(M_{\rm c}+M_{\rm e})$.}
    \label{fig:gp}
\end{figure}

We model a giant planet as a spherical object that consists of an incompressible core of radius $R_{\rm c}$ and a compressible envelope that extends from $R_{\rm c}$ to $R$, outside of which the density is zero (see Figure \ref{fig:gp}). As we noted in Section \ref{sec:introduction}, the processes of ionization and molecular dissociation that occur within the envelope imply that energy can be transferred from one species to another. Instead of modeling these processes explicitly, we treat the gas as adiabatic with an \emph{effective} 
adiabatic index that -- owing to these processes -- can be less than $4/3$. In this way, we do not account for the time dependence that accompanies these non-ideal processes, but instead gain some understanding as to the combined effects of the smaller adiabatic index and the presence of the incompressible core on the hydrodynamic stability of the envelope.
\subsection{Fluid equations}
\label{sec:equations}
With this set of assumptions, the hydrodynamical evolution of the envelope is governed by the continuity of mass, radial momentum, and entropy, the conservation laws for which read respectively: 
\begin{equation}
    \frac{\partial \rho}{\partial t}+\frac{1}{r^2}\frac{\partial}{\partial r}\left(r^2\rho v\right) = 0, \label{eq:continuity}
\end{equation}
\begin{equation}
    \frac{\partial v}{\partial t}+v\frac{\partial v}{\partial r}+\frac{1}{\rho}\frac{\partial p}{\partial r} = -\frac{GM_{\rm c}}{r^2}-\frac{GM}{r^2}, \label{eq:radialmomentum}
\end{equation}
\begin{equation}
    \frac{\partial}{\partial t}\ln\left(\frac{p}{\rho^{\gamma}}\right)+v\frac{\partial}{\partial r}\ln\left(\frac{p}{\rho^{\gamma}}\right) = 0. \label{eq:entropy}
\end{equation}
Here $\rho$ is the gas density within the envelope, $v$ is the radial velocity, $r$ is spherical radius from the origin, $p$ is the gas pressure, and $\gamma$ is the adiabatic index. On the right-hand side of Equation \eqref{eq:radialmomentum} we accounted for the gravitational presence of the core that has mass $M_{\rm c}$, and  

\begin{equation}
    M = \int_{r_{\rm c}}^{r}4\pi r^2\rho \,dr \label{eq:mofr}
\end{equation}
is the total atmospheric mass contained within radius $r$. It is mathematically convenient to write the continuity equation \eqref{eq:continuity} in terms of $M$, which is (upon using Equation \eqref{eq:mofr} to write $\rho$ in terms of $M$):

\begin{equation}
    \frac{\partial M}{\partial t}+v\frac{\partial M}{\partial r} = 0. \label{eq:contM}
\end{equation}
While we did not write it explicitly, all of the fluid variables are functions of both spherical radius $r$ and time $t$; we do not consider angular perturbations here.

We anticipate finding solutions to Equations \eqref{eq:radialmomentum} -- \eqref{eq:contM} that are hydrostatic, such that the radial velocity $v$ and time dependence are identically zero, on top of which we impose perturbations that induce radial motion and temporal evolution. The solutions for the fluid variables are therefore characterized by the existence of a surface $R(t)$ that, when the perturbations are small and the motions are subsonic, is approximately independent of time and separates the planetary interior from the ambient gas (which is assumed to have a negligible impact on the dynamics of the envelope). 
In our model we neglect accretion from the surrounding, protoplanetary disc, and hence the total mass of the planet $M_{\rm c}+M_{\rm e}$ is conserved, 
where $M_{\rm c}$ is the core mass and $M_{\rm e}$ is the total mass of the gaseous envelope. The characteristic speed within the envelope is governed by the sound speed $c_{\rm s}$, which -- owing to the approximately hydrostatic nature of the envelope -- is comparable to the freefall speed, or $c_{\rm s} \simeq \sqrt{G\left(M_{\rm c}+M_{\rm e}\right)/R}$. The characteristic timescale that parameterizes the temporal evolution of the planet is then the sound crossing time, being $R/c_{\rm s} \simeq R^{3/2}/\sqrt{G\left(M_{\rm c}+M_{\rm e}\right)}$.

Given these considerations, we analyze the fluid equations in terms of the following (dimensionless) space-like and time-like variables: 

  \begin{equation}
      \xi= \frac{r}{R(t)},
      \label{eq:coordinatetransform}
  \end{equation}
\begin{equation}
 d\tau=\frac{1}{R} \left(\sqrt{\frac{G(M_c+M_e)}{R}}\right)dt.
 \end{equation}
 We further non-dimensionalize the mass coordinate by defining 
 
 \begin{equation}
     M = M_{\rm e}m(\xi,\tau),
 \end{equation}
 from which it follows that
 
 \begin{equation} \label{eq:density}
     \rho = \frac{M_{\rm e}}{4\pi R^3}\frac{1}{\xi^2}\frac{\partial m}{\partial \xi} \equiv \frac{M_{\rm e}}{4\pi R^3}g(\xi,\tau), 
 \end{equation}
 where
  \begin{equation}
  g(\xi, \tau)= \frac{1}{\xi^2} \frac{\partial m}{\partial \xi}. \label{geq}
 \end{equation}
Similarly, we non-dimensionalize the pressure $p$ and the radial velocity $v$ by introducing the functions $h$ and $f$, defined via

  \begin{equation}
  \begin{split}
 &p=\left(\frac{GM_e(M_c+M_e)}{4 \pi R^4}\right) h(\xi, \tau), \\
 &v=\left(\sqrt{\frac{G(M_c+M_e)}{R}}\right) f(\xi, \tau).
 \end{split}
 \label{eq:coordinatetransformend}
 \end{equation}

We can now write the hydrodynamic equations \eqref{eq:radialmomentum}, \eqref{eq:entropy} and \eqref{eq:contM} in dimensionless form by introducing these coordinate transformations. To further simplify the resulting equations, we maintain that the fluid motions are subsonic, such that the dimensionless function $f$ satisfies $f \ll 1$ and $\partial R/\partial t \ll c_{\rm s}$. In this limit, we can also linearize the hydrodynamic equations in the assumed-small quantities $f$ and $1/c_{\rm s}\times\partial R/\partial t$, such that we keep only leading-order terms in these quantities and omit any non-linear contributions. Doing so yields the following three dimensionless,  fluid equations:
\begin{equation}
 \frac{\partial m}{\partial \tau}+\left(f- 
 V\xi\right)\frac{\partial m}{\partial \xi}=0,
 \label{eq:nvconm}
 \end{equation}
    \begin{equation}
 \frac{\partial f}{\partial \tau}+\frac{1}{g}\frac{\partial h}{\partial \xi}=-\frac{\mu}{\xi^2} - \left(1-\mu\right) \frac{m}{\xi ^2},
 \label{eq:nvmomentum} 
 \end{equation} 
 \begin{equation}
  \frac{\partial}{\partial \tau}\ln\left(\frac{h}{g^{\gamma}}\right)+V\left(3\gamma-4\right)+\left(f-V\xi\right)\frac{\partial}{\partial \xi}\ln\left(\frac{h}{g^{\gamma}}\right) = 0. \label{eq:nentropy1}
 \end{equation} 
Here we defined 

\begin{equation}
    \mu \equiv \frac{M_{\rm c}}{M_{\rm c}+M_{\rm e}}    
\end{equation}
as the ratio of the mass of the core to the total mass of the planet, and 

\begin{equation}
    V \equiv \frac{\partial R}{\partial t}\left(\frac{G\left(M_{\rm c}+M_{\rm e}\right)}{R}\right)^{-1/2}
\end{equation}
is the ratio of the velocity of the surface of the envelope to the escape speed. 

In the absence of conduction, the surface of the planet is a contact discontinuity that separates the planetary interior from the ambient gas, across which the pressure and fluid velocity (in the comoving frame of the contact discontinuity) are continuous. We further reduce the parameter space of our solutions by assuming that the ambient gas does not play a dominant role in providing further pressure confinement of the planetary envelope, and therefore the ambient pressure is set to zero. With these assumptions, the boundary condition on the fluid velocity at the surface is 

\begin{equation}
 v(\xi=1, \tau)=\frac{\partial R}{\partial t} \quad \Rightarrow \quad f(\xi=1, \tau)= V. \label{bcvel}%V \left(\sqrt{\frac{G(M_c+M_e)}{R}}\right)^{-1} 
\end{equation}
In addition, the pressure perturbation at the surface must vanish to smoothly match onto the surrounding medium, and if the planet is subject to only radial perturbations, the fluid velocity at the core radius must be zero. In the next two sections, we seek solutions to the above set of equations that satisfy these boundary conditions by treating the envelope as a hydrostatic medium with small, time-dependent perturbations, where ``small'' implies that the radial velocity is subsonic. As such, we write

\begin{equation}
    h(\xi,\tau) = h_0(\xi)+h_1(\xi,\tau),
\end{equation}
\begin{equation}
    g(\xi,\tau) = g_0(\xi)+g_1(\xi,\tau),
\end{equation}
\begin{equation}
    f(\xi,\tau) = f_1(\xi,\tau)
\end{equation}
with subscript-0 quantities representing the unperturbed, hydrostatic solutions and subscript-1 quantities the perturbations. Note that, because the background upon which we impose such perturbations is hydrostatic, there is no zeroth-order velocity (i.e., $f_0$ is absent from the above expressions). Below we will also work with the functions $m_0$ and $m_1$, which are directly related to $g_0$ and $g_1$ via Equation \eqref{geq}.

\subsection{Hydrostatic Solutions}  \label{sec:static}
The hydrostatic solutions to Equations \eqref{eq:nvconm} - \eqref{eq:nentropy1} are independent of time and possess no radial velocity, meaning that $V \equiv 0$ and $f_0 = 0$. In this case, the continuity and entropy equations are trivially satisfied, and the radial momentum equation yields the dimensionless equation of hydrostatic balance that relate the subscript-0 functions: 

\begin{equation}
\frac{1}{g_0}\diffp{h_0}{\xi}=-\mu\frac{1}{\xi^2} - (1-\mu) \frac{m_0}{\xi ^2},
 \end{equation}
To close this system, we need a relationship between the hydrostatic pressure and the density, for which we adopt a polytropic equation of state and let 

\begin{equation}
    h_0 = K_0g_0^{\gamma},
\end{equation}
where $K_0$ is the dimensionless specific entropy of the atmospheric gas.
The equation for $m_0$ is then
 \begin{equation} 
\frac{\gamma}{\gamma-1} K_0 \diffp{}{\xi} \left[\left(\frac{1}{\xi^2}\diffp{m_0}{\xi}\right)^{\gamma-1}\right]=-\frac{\mu}{\xi^2}-(1-\mu)\frac{ m_0}{\xi^2}. \label{eq:laneemden}
\end{equation} 
Note that, if the first term on the right-hand side of this equation were absent, this would just be the Lane-Emden equation written in terms of the enclosed mass $m_0$. When $\mu \neq0$, the first term represents the gravitational contribution from the core. 

We can numerically solve Equation \eqref{eq:laneemden} by following the procedure outlined in \citet{coughlin20}: the solutions to this equation must satisfy the requirement that the total mass of the envelope be equal to $M_e$, so that $m_0(1)=1$. We can therefore expand the function $m_0$ in powers of $(1-\xi)$ and equate like powers on the left and right sides of Equation \eqref{eq:laneemden}; this results in the following, leading-order approximation to $m_0$ near the surface of the planet:

\begin{multline} 
m_0(\xi)= 1-\frac{\gamma-1}{\gamma}\left(\frac{\gamma-1}{\gamma} \frac{1}{K_0}\right)^{\left(\frac{1}{\gamma-1}\right)}(1-\xi)^{\left(\frac{\gamma}{\gamma-1}\right)} \\ 
\times \left[1-\left(\frac{2\gamma-3}{2\gamma-1}\right)(1-\xi)\right]. \label{sollaneemden}
\end{multline} 
This expression can be used to evaluate $m_0(\xi \simeq 1)$ and $dm_0/d\xi(\xi \simeq 1)$, which can be used as boundary conditions to integrate Equation \eqref{eq:laneemden} inward for a given, dimensionless specific entropy $K_0$. 
When the polytropic envelope is assumed to extend to the geometric center of the planet, $K_0$ is fixed by requiring that the enclosed mass equal zero at the origin. For a given $\gamma$, there is then a unique value of $K_0$ for which the density terminates at the surface ($\xi \simeq 1$) and the enclosed mass equals zero at the origin, which is just the usual solution to the Lane-Emden equation (e.g., \citealt{Liu96, Chavanis}). This value of $K_0$ can be determined iteratively through a brute-force, trial-and-error method in which one successively changes the value of $K_0$ until the boundary condition near the origin is satisfied (see the discussion in Section 2 of \citealt{coughlin20}). 

\begin{figure*} 
    \centering
    \includegraphics[width=0.325\textwidth]{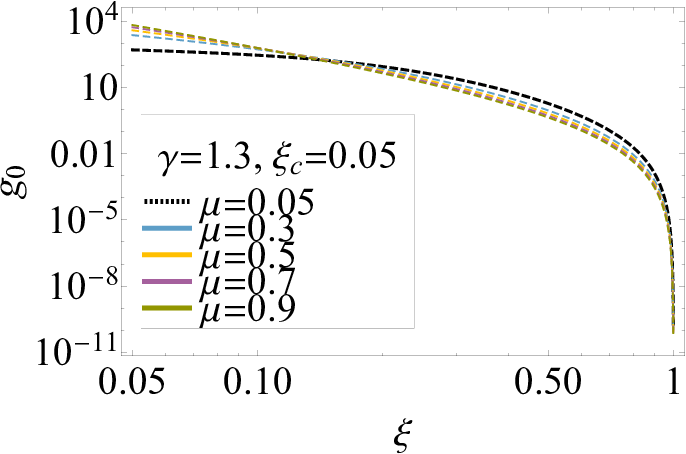}
    \includegraphics[width=0.325\textwidth]{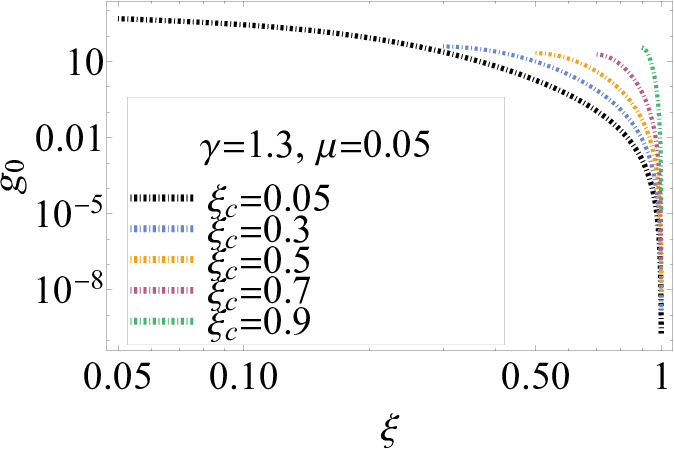}
    \includegraphics[width=0.325\textwidth]{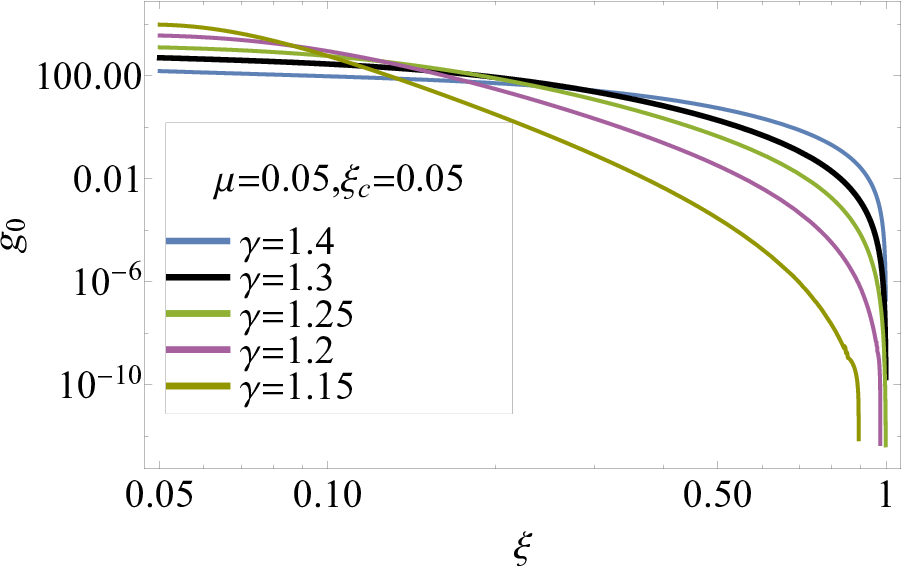}
    \caption{Left: The unperturbed and dimensionless density $g_0$ as a function of radius for the relative core masses $\mu$ shown in the legend and a polytropic index $\gamma = 1.3$ and a core radius of $\xi_{\rm c} = 0.05$. 
    Middle: Identical to the left panel but with fixed $\gamma = 1.3$ and $\mu = 0.05$ and the inner radii shown in the legend. 
    Right: Identical to the left panel but with fixed $\mu = 0.05$, fixed $\xi_{\rm c} = 0.05$, and variable adiabatic indices as shown in the legend. 
    }
 \label{fig:hs}
\end{figure*}
For our application here, however, the existence of an incompressible core of relative mass $\mu$ modifies the solution from the usual, Lane-Emden function. In particular, if we let the core have some associated, relative radius $\xi_{\rm c}$ (i.e., $\xi_{\rm c}$ is the radius of the core, $R_{\rm c}$, divided by the total radius of the planet $R$), then the total enclosed mass of the envelope must instead be zero at $\xi_{\rm c}$. In addition to the relative mass of the core $\mu$, there is thus an additional, free parameter contained in these solutions, which is the inner radius $\xi_{\rm c}$ (presumed greater than zero) at which the core terminates and the envelope begins. The solutions to Equation \eqref{eq:laneemden} are therefore determined by three parameters, being the core mass $\mu$, core radius $\xi_{\rm c}$, and adiabatic index $\gamma$; once these three parameters are fixed, the value of $K_0$ that satisfies the inner boundary condition ($m_0(\xi_{\rm c}) = 0$) can be determined by -- as for the standard Lane-Emden equation described above -- iteratively looping over $K_0$ in a trial-and-error manner.

\begin{figure*} 
    \centering
     \includegraphics[width=0.325\textwidth ]{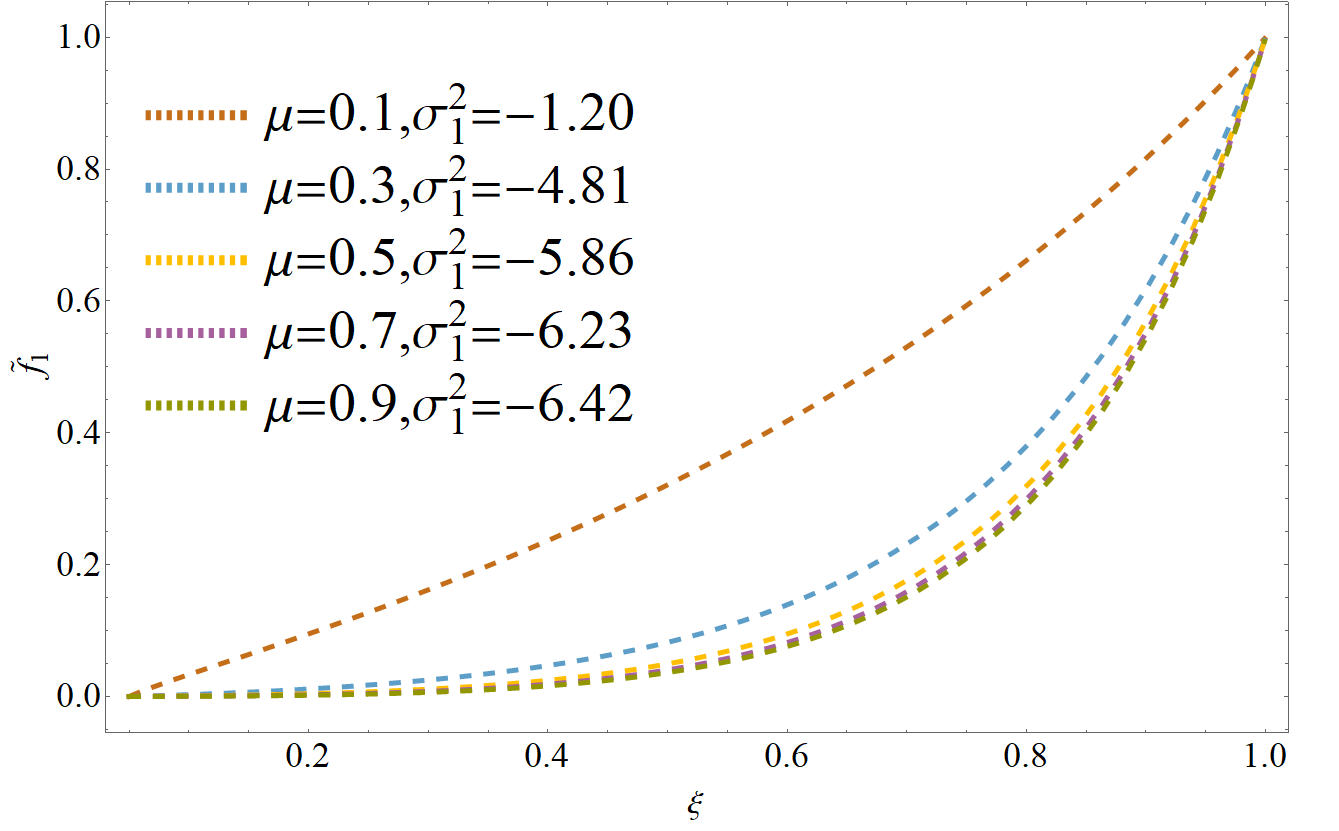}
    \includegraphics[width=0.325\textwidth]{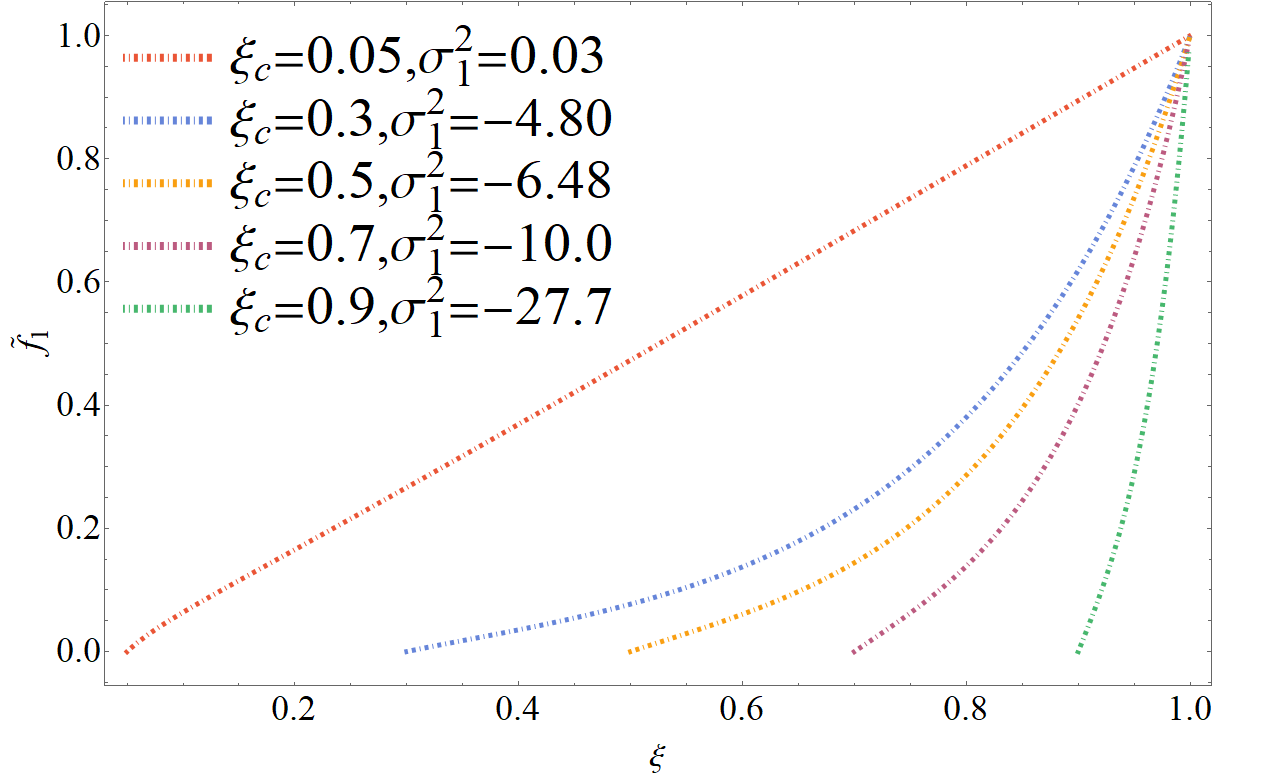}
    \includegraphics[width=0.325\textwidth]{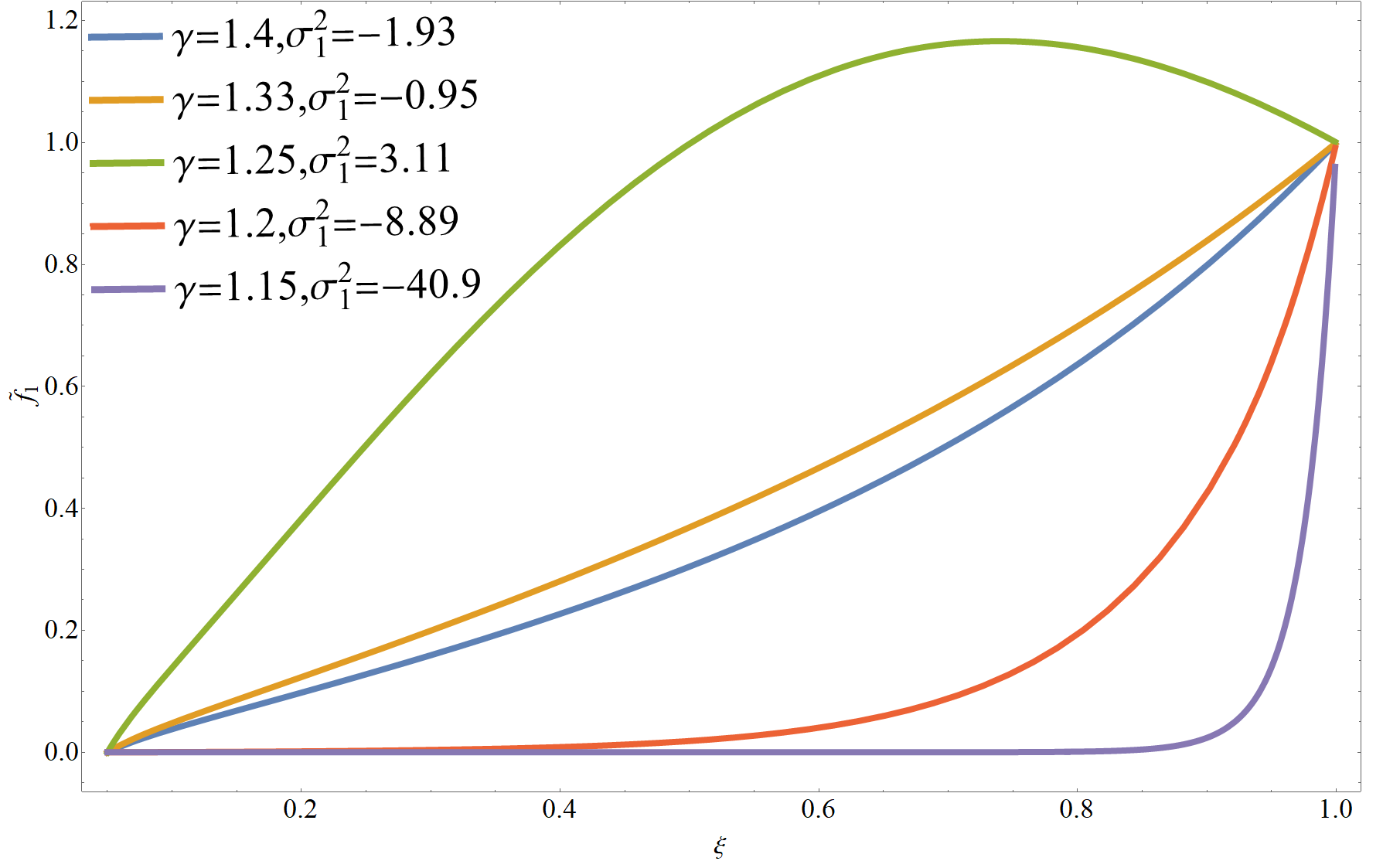}
    \caption{Top-left panel: The lowest-order eigenfunctions as a function of the dimensionless radius for the polytropic index $\gamma=1.3$ and core radius $\xi_c=0.05$ for different values of the fractional core mass $\mu$ shown in the legend. 
   }    
   \label{fig:eigenfuntions}
\end{figure*}

\begin{figure*} 
    \centering
     \includegraphics[width=0.495\textwidth]{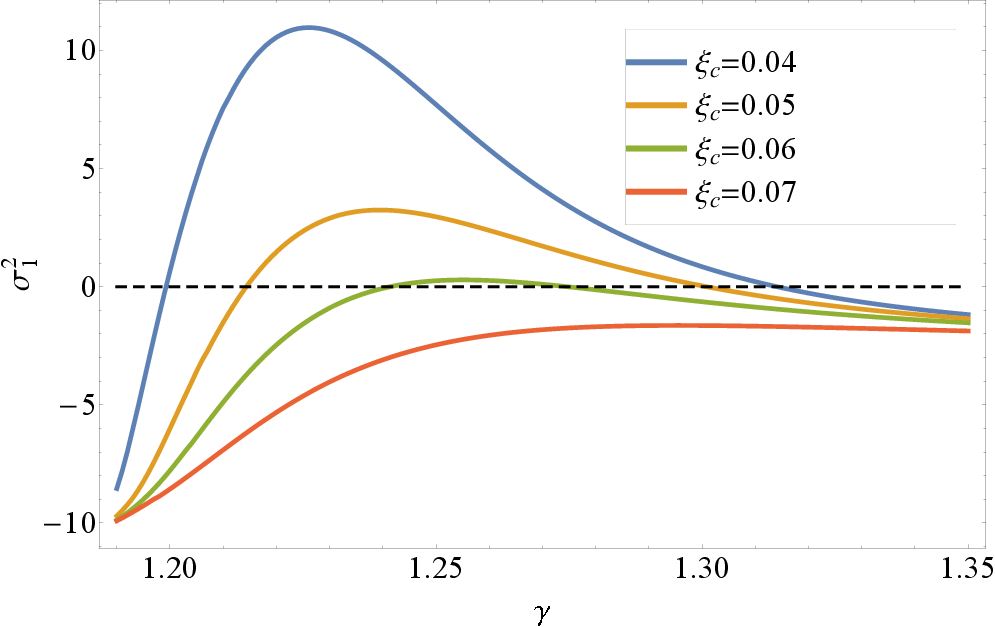}
      \includegraphics[width=0.495\textwidth]{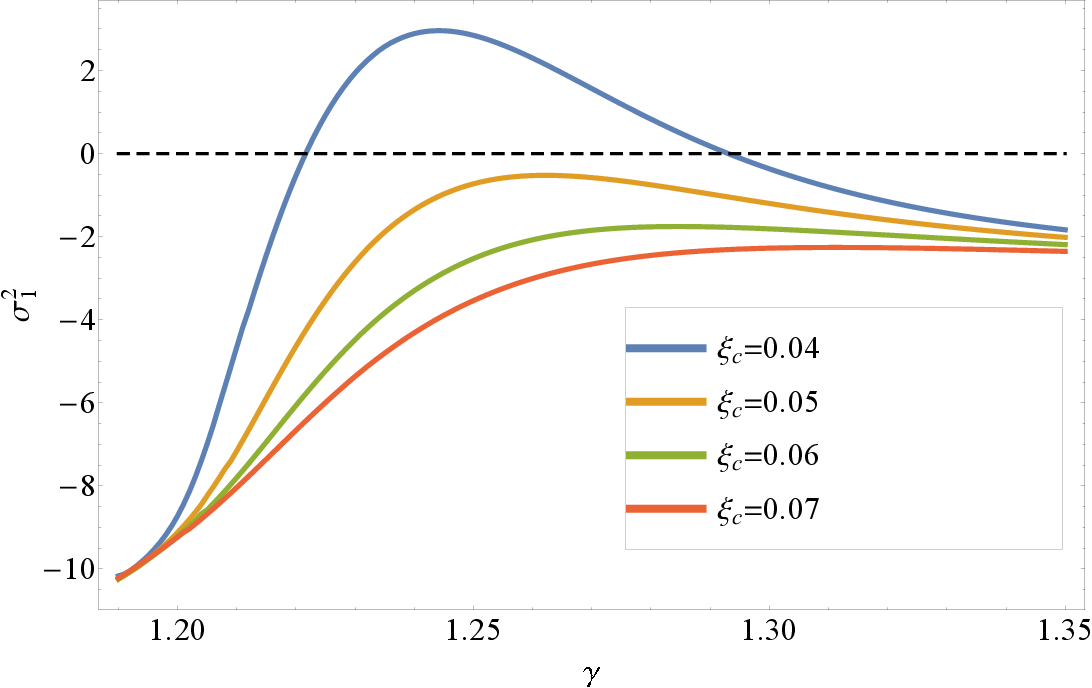}
    \caption{The square of the lowest-order eigenvalue as a function of the adiabatic index, $\gamma$. The left panel corresponds to a core mass of $\mu = 0.05$, while in the right panel we set $\mu = 0.1$. 
    }
     \label{fig:eigenvalsqr}
\end{figure*}
In principle, the solutions for the envelope extend asymptotically close to $\xi = 1$ where the density is exactly zero and the mass satisfies $m \equiv 1$. However, the envelope becomes extremely tenuous near the surface once $\gamma$ nears unity -- the case of interest here -- owing to the fact that the density declines approximately as $\propto \left(1-\xi\right)^{1/\left(\gamma-1\right)}$ (e.g., for $\gamma = 1.1$, the density reaches $\sim 10^{-10}$ at $\xi = 0.9$). In practice and to avoid any numerical artifacts associated with initializing the integration of Equation \eqref{eq:laneemden} too close to the surface, we integrate Equation \eqref{eq:laneemden} from the location where leading-order solution for the unperturbed density (i.e., using Equation \ref{sollaneemden}) satisfies $g_0 = 10^{-8}$, and outside of this location we replace the numerically obtained solution with the leading-order, series expansion given in Equation \eqref{sollaneemden}. We have verified that increasing or decreasing the threshold, lower limit on the density that sets the outer boundary does not noticeably change the hydrostatic solution or the eigenmodes (see below). 
Examples of the hydrostatic solutions are shown in Figure (\ref{fig:hs}). To isolate the impact of the relative core mass ($\mu$), radius $(\xi_{\rm c}$), and polytropic index ($\gamma$) on the hydrostatic solutions, we fix all but one parameter for each of the plots. The left panel of Figure (\ref{fig:hs}) shows five different solutions for the dimensionless density $g_0$ (on a logarithmic scale) as a function of the normalized spherical radius variable $\xi$, obtained by setting the adiabatic index $\gamma=1.3$, the dimensionless core radius $\xi_c=0.05$ and varying the fractional core mass $\mu$, successively for each curve. Comparing these five curves (as we move from $\mu=0.9$ to $\mu=0.05$) we observe two contrasting behaviors: the configuration that receives the highest contribution from core to its total mass $(\mu=0.9)$, is the densest configuration near the core $(\xi \approx \xi_c)$ and as we move toward the outer edge $(\xi \approx 1)$ its density ranks to be the least dense one. This is due to the fact that the gravitational field is more centrally concentrated for a larger core mass, and all the material piles up near the inner boundary. 

In the middle panel of this figure we show solutions obtained by varying the size of the core (successively as we move from one curve to the next) while maintaining $\gamma=1.3$ and $\mu=0.05$. We see that the envelope is denser on average, for a more voluminous core. This trend is expected because as the core grows in size, it leaves less space to be occupied by the envelope that -- for the same $\mu$ -- contains the same mass, resulting in an increase in density. 

In the right panel of this figure we vary the adiabatic index $\gamma$ while maintaining a fixed dimensionless core radius $\xi_c = 0.05$ and fractional core mass $\mu = 0.05$. We see the same contrasting behavior between regions near the inner and outer boundaries. We notice the most compressible envelope (the one with the smallest adiabatic index $\gamma=1.15$), ranks to be the densest, closer to the core $(\xi \approx \xi_c)$, and the least dense near the outer edge of the planet $(\xi \approx 1)$. This behavior occurs because the increased compressibility of the gas causes more matter to pile up near the core. 

The non-dimensionalized equations we derived demonstrate that hydrostatic solutions to the fluid equations in the absence of an entropy gradient are manifestly scale-free; this is exploited in the standard analysis of the Lane-Emden equation by defining

\begin{equation}
    \alpha^2 = \frac{\gamma}{\gamma-1}\frac{K\rho_{\rm c}^{\gamma-1}}{4\pi G\rho_{\rm c}},
\end{equation}
where $K$ is the %dimensional entropy 
adiabatic constant related to the central pressure $p_c$ and density $\rho_{\rm c}$ as $\frac{p_c}{\rho_c^{\gamma}}$. The surface of the planet then occurs at some fixed number times $\alpha$. However, it is a choice to work in coordinates normalized by $\alpha$ and the central properties of the planet, and we can instead -- as we have done here -- choose to work with a radial coordinate that is relative to the surface and the average properties of the planet. In this case the \emph{dimensionless} entropy variable $K_0$ that ensures the regularity of the solutions at the center of the planet is a function of the adiabatic index, and here it is also a function of the core mass and radius. Owing to the self-similarity of the equations, the same solution is valid for any choice of physical planet radius $R$ provided that the radius is scaled by $R$, the pressure is scaled by $M/R^{-4}$, and the density by $M/R^{-3}$. 

\section{Perturbation Analysis}  
\label{sec:perturbations}
\subsection{Eigenmode Equations}
With the polytropic solution as the unperturbed state, the perturbation equations are
\begin{equation} \label{eq:perturbedconm}
\diffp{m_1}{\tau}= \left(V\xi-f_1\right)\diffp{m_0}{\xi},
\end{equation}
\begin{equation} \label{eq:perturbedlinearmomentum}
\diffp{f_1}{\tau} -\frac{g_1}{g_0^2}\diffp{h_0}{\xi}+\frac{1}{g_0}\diffp{h_1}{\xi}= -(1-\mu) \frac{m_1}{\xi ^2}, 
\end{equation}
\begin{equation} \label{eq:perturbedentropy}
    \frac{\partial }{\partial \tau} \left[\frac{h_1}{h_0}-\frac{\gamma g_1}{g_0}\right]-V(4-3\gamma) = 0. 
\end{equation} 
Note that, in the last of these expressions, the Brunt-Vaisala frequency is identically zero -- and the solutions are therefore buoyantly neutral -- because of the isentropic nature of the envelope.
We can now take the Laplace transform of Equations \eqref{eq:perturbedconm}--\eqref{eq:perturbedentropy}, where the Laplace transform of $f_1$ is 
\begin{equation}
 \tilde f_1(\xi,\sigma)= \int_{-\infty}^{\infty}f_1(\xi,\sigma) \exp{(-\sigma \tau)} \, d\tau
\end{equation}
and similarly for all other fluid variables. We can understand the linear response of the envelope to a given perturbation by letting there be, for example, an initial, non-zero velocity with some underlying radial dependence; this function then appears on the right-hand side of the Laplace-transform of Equation \eqref{eq:perturbedlinearmomentum} and serves to initialize the motion of the gas. The resulting response is then able to be written as a sum over the \emph{eigenmodes} of the set of Laplace-transformed equations, where the eigenmodes are solutions to this set of equations that possess \emph{eigenvalues}. The eigenvalues are complex numbers that we denote $\sigma_{\rm n}$ where the perturbations to the fluid variables diverge as simple poles in the complex plane. Importantly, even though the coefficients in such an eigenmode expansion depend on the nature of the initial perturbation, the fact that the fluid variables diverge at the eigenvalues implies that the eigenmodes themselves do not (see \citet{coughlin20} for explicit expressions for the coefficents). Defining $\tilde{f}_{\rm n} = \tilde{f}_1/\tilde{V}$ as the ratio of the perturbation to the fluid velocity to the perturbation to the surface velocity and taking the limit as $\sigma \rightarrow \sigma_{\rm n}$, a single, second-order equation for $\tilde{f}_{\rm n}$ can be derived by combining the Laplace-transformed equations; the equation is 
\begin{multline} \label{eq:eigenmode}
%\begin{split}
\sigma_{\rm n}^2 \tilde{f}_{\rm n} + \frac{\partial}{\partial \xi}\left[{\frac{\gamma h_0}{g_0}\frac{1}{g_0 \xi^2} \frac{\partial}{\partial \xi}{\left[\left(\xi -\tilde{f}_{\rm n}\right)\diffp{m_0}{\xi}\right]}}\right]
+(4-3 \gamma)\frac{1}{g_0} \diffp{h_0}{\xi} \\ +
\frac{1-\mu}{\xi^2}(\xi-\tilde{f}_{\rm n})\diffp{m_0}{\xi} = 0.
\end{multline}
The fluid velocity must be continuous across the surface of the planet, which gives the first boundary condition on the eigenfunction (cf.~Equation \ref{bcvel})

\begin{equation} \label{eq:bc1}
    \tilde{f}_{\rm n}(\xi = 1) = 1. 
\end{equation} 
We also demand that the solutions for the eigenmodes be non-trivial and expandable about the surface; taking the leading-order terms in the series expansion of Equation \eqref{eq:eigenmode} then shows that the derivative of $\tilde{f}_{\rm n}$ satisfies

\begin{equation} \label{eq:bc2}
    \frac{\partial \tilde{f}_{\rm n}}{\partial \xi}\bigg{|}_{\xi = 1} = \frac{1}{\gamma}\left(\sigma_{\rm n}^2+2\left(\gamma-2\right)\right). 
\end{equation}
Finally, the fluid velocity must be equal to zero at $\xi_{\rm c}$ owing to the incompressible nature of the core, which gives the additional boundary condition

\begin{equation} \label{eq:bc3}
    \tilde{f}_{\rm n}(\xi_{\rm c}) = 0.
\end{equation}
This third boundary condition determines the eigenvalues $\sigma_{\rm n}$, as we start with some initial guess for the eigenvalue, integrate equation \eqref{eq:eigenmode} inward from a point near the surface ($\xi \simeq 1$) using boundary conditions \eqref{eq:bc1} and \eqref{eq:bc2} with this guess, and determine the value of $\tilde{f}_{\rm n}(\xi_{\rm c})$. We then perturb the guess for $\sigma_{\rm n}$, calculate the new function and the corresponding residual of $\tilde{f}_{\rm n}(\xi_{\rm c})$, and continue to iterate on the value of $\sigma_{\rm n}$ until we satisfy the third boundary condition $\tilde{f}_{\rm n}(\xi_{\rm c}) = 0$. The set of eigenvalues $\sigma_{\rm n}$ then delimit the solutions that satisfy the boundary conditions near the surface of the planet and at the inner core. Since Equation \eqref{eq:eigenmode} is in the form of a Hermitian operator equation (e.g., \citealt{hansen04}), the eigenvalue squares $\sigma_{\rm n}^2$, are purely real, and the corresponding solutions vary as $\propto \exp\left(\sigma\tau\right)$. Thus, if $\sigma_{\rm n}^2 < 0$, the solutions are stable and oscillate in time, whereas if $\sigma_{\rm n}^2 > 0$ the envelope is unstable and small perturbations grow exponentially rapidly with time. The lowest-order mode $\tilde{f}_1$ has no zero crossings, and each higher-order mode has one more zero crossing than the previous one.
\begin{table*}
    \centering
   \begin{tabular}{|c|c|c|c|c|c|c|c|c|}
 \hline 

\multicolumn{3}{|c|}{$\gamma=1.3, \, \xi_c=0.05$} &  \multicolumn{3}{|c|}{$\gamma=1.3, \, \mu=0.05$} & 
\multicolumn{3}{|c|}{$\mu=0.05,\, \xi_{\rm c}=0.05$} \\
\hline
 
$\mu$ & $K_0$ & $\sigma_1^2$ & $\xi_{\rm c}$ & $K_0$ & $\sigma_1^2$ & $\gamma$ & $K_0$ & $\sigma_1^2$ \\
\hline
0.1 & 0.202 & -1.20 &	0.05 & 0.190 & 0.03 & 1.4 &	0.136 & -1.93 \\
\hline
0.3 & 0.238 & -4.81  & 0.3 &	0.108 &	-4.80 & 1.33 & 0.168 & -0.95  \\
\hline
0.5 & 0.263 & -5.86 & 0.5 &	0.0633 &	-6.48 & 1.25 & 0.246 & 3.11 \\

\hline
0.7 & 0.282 & -6.23  & 0.7 & 0.0303 &	-10.0 & 1.2 & 0.333 & -8.89  \\

\hline
0.9 & 0.298 & -6.42  & 0.9 &	0.00683 &	-27.7  & 1.15 & 0.439 & -40.9   \\
\hline 
\end{tabular}
    \caption{The square of the lowest-order eigenvalue $\sigma_{1}^2$ and the auxiliary variable $K_0$ %\sout{has been tabulated} 
    for each of the configurations shown in Figure (\ref{fig:hs}).}
    \label{tab:my_label}
\end{table*}

\begin{figure*}
    \centering
     \includegraphics[width=0.495\textwidth]{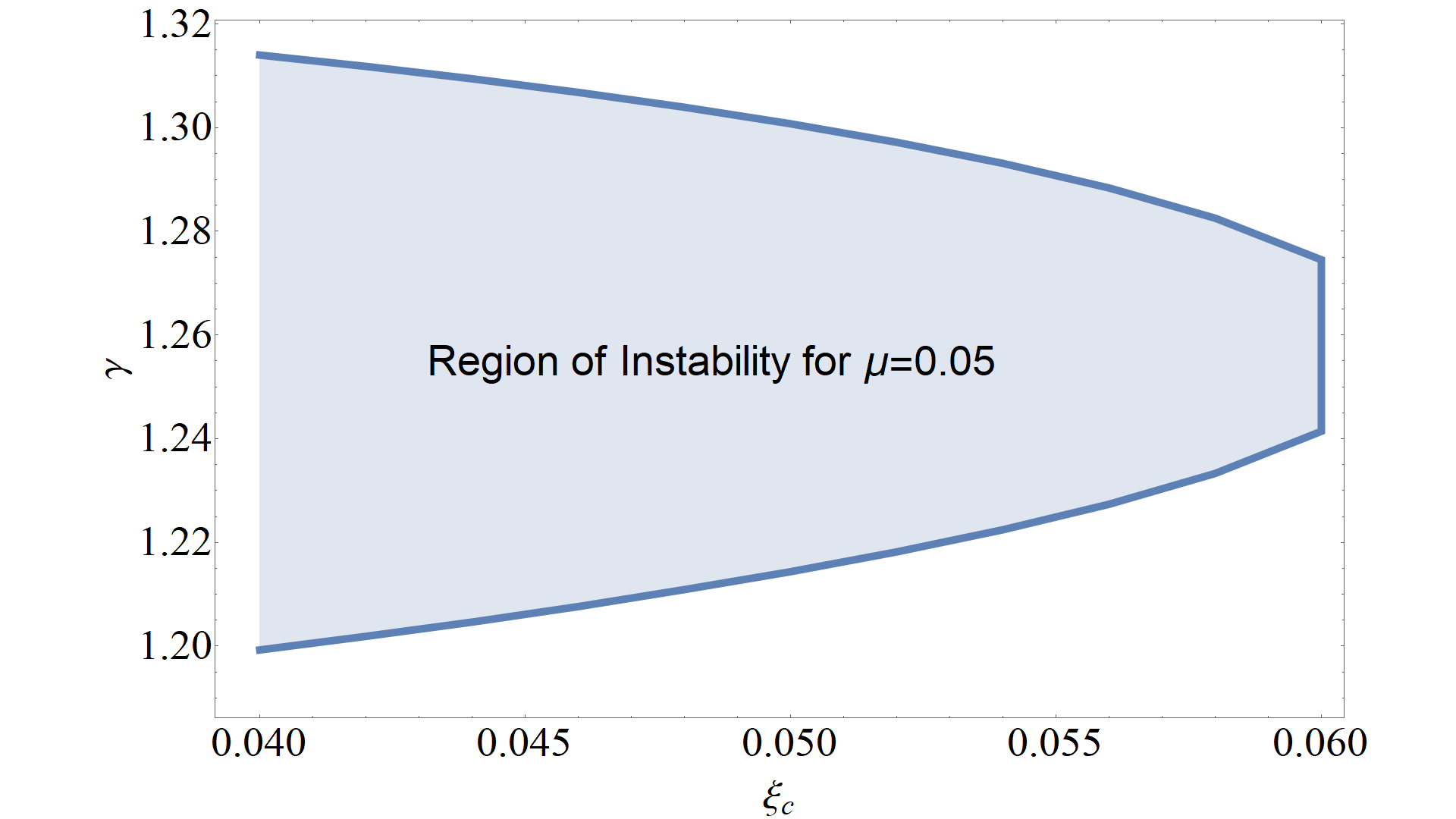}
     \includegraphics[width=0.495\textwidth]{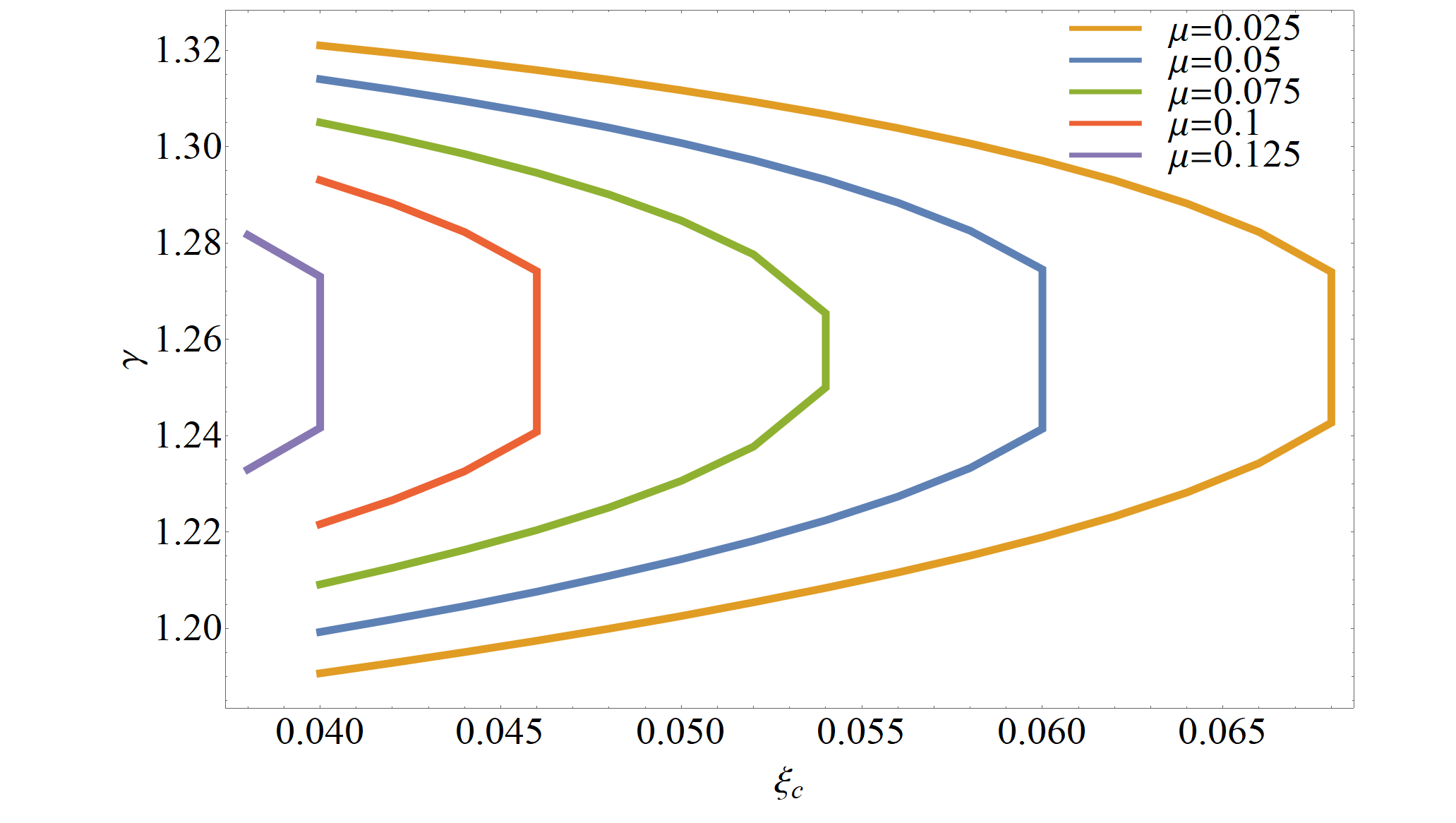}
    \caption{The curve that delimits where $\sigma_1^2 = 0$ in the $\gamma-\xi_c$ plane that separates the stable and unstable behavior. The left panel fixes $\mu = 0.05$ and the shaded region denotes where the solutions are unstable, while the right panel varies the core mass.
    }
    \label{fig:Omega_0}
\end{figure*}

\begin{figure} 
    \centering
      \includegraphics[width=0.4\textwidth]{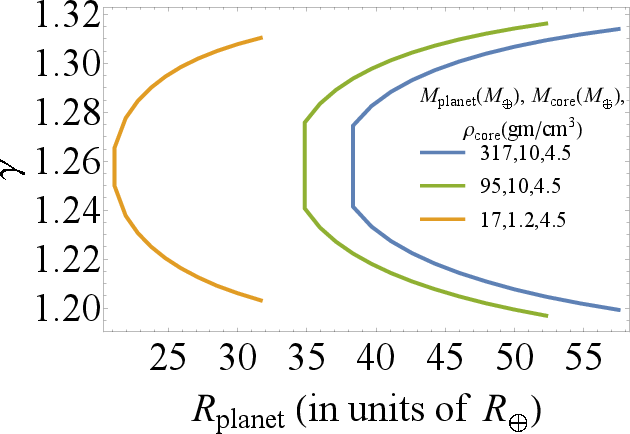}
    \caption{The $\sigma_1^2 = 0$ curve separating stable and unstable behavior. We have simulated three model planetary configurations having similar properties to the gas-giants in our solar system:
    Jupiter(Blue), Saturn(Green), Neptune(Orange). The model parameters are shown in legends.
    }
     \label{fig:Omega_0_astr}
\end{figure}

As for the hydrostatic solutions, we set the outer boundary (at which we evaluate the boundary conditions \ref{eq:bc1} and \ref{eq:bc2}) at the location within the hydrostatic envelope at which the dimensionless, unperturbed density satisfies $g_0 = 10^{-8}$. We have verified that changing this small parameter by an order of magnitude (increasing or decreasing) has no effect on the eigenvalues or the eigenfunctions. To further mitigate any numerical errors, in practice in the eigenmode equation \eqref{eq:eigenmode} we replace any derivatives of $m_0$ higher than the first with the lower-order values that result from the equation of hydrostatic equilibrium \eqref{eq:laneemden}.

\subsection{Solutions}
In 
Figure \ref{fig:eigenfuntions} we present the lowest-order eigenfunction 
for the hydrostatic solutions presented in Figure \ref{fig:hs}; the eigenvalues that characterize these solutions are given in Table \ref{tab:my_label} (along with other properties of the hydrostatic envelope). We see from this figure that, as the eigenvalue appropriate to the lowest-order mode becomes increasingly negative, the eigenfunction decreases more rapidly from the surface of the envelope (which can be seen directly from the boundary condition on the derivative of $\tilde{f}_{\rm n}$ in Equation \ref{eq:bc2}). This feature is a familiar property of the $p$-modes of linearly stable stars (e.g., \citealt{hansen04}), and illustrates that the majority of the power of these modes resides in the outer layers of the planetary envelope. Thus, a solution that is characterized by only very negative eigenvalues responds to a homologous initial velocity perturbation (i.e., such that the initial velocity profile is $\propto \xi$) by oscillating violently and stochastically -- albeit stably and sinusoidally in time -- in its outermost extremities (see the end of Section 3.1 of \citealt{coughlin20} for further discussion).

On the other hand, as the lowest-order eigenvalue becomes less negative, the eigenfunction is better approximated by a linear function of $\xi$ that extends from the core to the planet radius. Indeed, if $\gamma=4/3$ and the inner boundary extends to the origin, then an exact solution to Equation \eqref{eq:eigenmode} and the boundary conditions \eqref{eq:bc1}--\eqref{eq:bc3} is $\tilde{f_n}=\xi$ with $\sigma_n^2=0$, which is just the familiar Chandrasekhar limit. Thus, a linear initial velocity perturbation for this specific case retains its linear profile, and the entire envelope expands or contracts homologously with time. As the square of the eigenvalue becomes positive, the leading-order eigenfunction attains a relative maximum in the interior of the envelope. This feature illustrates that a contraction of the envelope results in the acceleration of the fluid in the interior of the planet, which increases the gravitational potential and further accelerates contraction, resulting in a runaway collapse of the gas that proceeds exponentially rapidly with time. In the opposite scenario of an initial expansion, the very slow decline of the pressure results in the outward acceleration of the envelope.
\subsection{Region of Instability} 
\label{sec:results}
It is clear from Figure \ref{fig:eigenfuntions} and Table \ref{tab:my_label} that, in addition to the singular case of the Chandrasekhar instability where $\sigma^2 = 0$ for $\xi_{\rm c} = \mu = 0$ and $\gamma = 4/3$, there is a region of parameter space within which the leading-order eigenvalue is positive and the envelope is dynamically unstable to radial perturbations. Qualitatively we expect that as the adiabatic index softens, the envelope should become more susceptible to instability, as the increase in the gas pressure is less pronounced (and less able to resist the increase in self-gravity) for an initial, radial contraction; an analogous argument for the existence of the instability exploits the less rapid decline in the pressure for an initial expansion of the envelope. Eventually, for a fixed core mass and radius, we expect this Chandrasekhar-like instability to arise once the adiabatic index crosses a sufficiently small, threshold value. 
To support this prediction, Figure \ref{fig:eigenvalsqr} shows the square of the leading-order eigenvalue as a function of $\gamma$ for a fixed core mass ($\mu = 0.05$ for the left panel and $\mu = 0.1$ for the right panel). Each curve in this figure is for a different value of the core radius $\xi_{\rm c}$, with the core radii appropriate to each curve given in the legend. We observe from the left panel that as we go to smaller values of $\gamma$, the leading-order eigenvalue becomes less negative, and provided that the core radius satisfies $\xi_{\rm c} \lesssim 0.06$, crosses $\sigma^2 = 0$ and the envelope becomes unstable. This zero crossing represents the generalization of the Chandrasekhar limit to a envelope that possesses a core (modeled as incompressible) in its interior.

For finite (i.e., non-zero) $\xi_{\rm c}$ and $\mu$, the $\gamma$ at which the envelope transitions from stable to unstable in going from large to small $\gamma$ is always less than $\gamma = 4/3$ (e.g., for $\xi_{\rm c} = \mu = 0.05$, from Figure \ref{fig:eigenvalsqr} the adiabatic index that separates stable and unstable envelopes is $\gamma \simeq 1.3$). This finding demonstrates that the effect of an incompressible core in the planetary interior is to \emph{stabilize} the envelope against the Chandrasekhar-like instability. This stabilizing influence arises from the fact that, because it is incompressible, there is no change in the gravitational potential associated with the core when we impose a perturbation to the planet. Thus, the destabilizing increase in self-gravity is always lessened in comparison to a core-less envelope, implying that the critical $\gamma$ that yields an unstable envelope must be less than the one that characterizes an envelope without a solid core.

Interestingly, Figure \ref{fig:eigenvalsqr} also shows that the leading-order eigenvalue does not increase monotonically as the adiabatic index decreases, and instead there is a \emph{most unstable envelope} that is characterized by a largest $\sigma_1^2$. As $\gamma$ continues to decrease below this most-unstable value, the square of the eigenvalue starts to decline, and crosses below zero to yield \emph{stable solutions} once $\gamma$ is sufficiently small. Moreover, for values of the core radius that are large enough, the envelope \emph{never becomes unstable}, even for very small values of $\gamma$, which disagrees with our naive expectation that sufficiently small adiabatic indices should give rise to an unstable envelope.

This behavior arises because as $\gamma$ decreases, all of the mass of the envelope becomes increasingly concentrated near the core, as can be seen from the right-most panel of Figure \ref{fig:hs}. However, since the fluid cannot penetrate the planetary core, any perturbation to such an envelope cannot result in a large change in the distribution of its mass and its corresponding gravitational potential (i.e., the eigenfunction satisfies $\tilde{f}_1(\xi_{\rm c}) = 0$, hence the vast majority of the mass is not displaced when acted upon by any perturbation, and the resulting change in the gravitational potential -- proportional to the change in the mass \emph{interior to radius $r$} -- is likewise small). Therefore, for sufficiently small $\gamma$, the envelope can effectively be considered a non-self-gravitating fluid with the entirety of the planetary mass located at the edge of the core; since the change in the self-gravity of this configuration must be effectively zero for any physical perturbation that does not displace the core, such an envelope cannot exhibit any Chandrasekhar-like instability.

The shaded region in the left-hand panel of Figure \ref{fig:Omega_0} shows the range of adiabatic indices within which the envelope is unstable as a function of the inner radius $\xi_{\rm c}$ when $\mu = 0.05$, and the curve that encloses this region delimits the combination of $\gamma$ and $\xi_{\rm c}$ -- for a fixed $\mu$ -- at which the eigenvalue crosses the $\sigma_1^2 = 0$ instability threshold. As the inner radius decreases, a larger range of adiabatic indices is unstable. The right-hand panel of this figure shows a set of such curves (within each of which the envelope is unstable) for a range of core masses. As the core mass increases, the envelope becomes stable to a larger range of inner radii, which arises from the fact that the perturbation to the self-gravity of the planet declines as more of the planetary mass is contributed by the incompressible core.
To show our results in explicit astrophysical units, we map out the instability region in Figure \ref{fig:Omega_0_astr} for planetary parameters (the parameters shown in legends) similar to those appropriate to the gas giants in our solar system. 

\subsection{Non-radial perturbations}
We consider the analysis of non-radial perturbations to be beyond the scope of this paper, but it may be useful to outline some of the considerations. To study angular perturbations with rigor, one can treat the solutions we obtained due to radial perturbations as the radial component of a total solution of the form $f(\xi,\theta,\phi,\tau)=f_{r}(\xi)Y_{\ell}^{m}(\theta,\phi)\exp(\sigma \tau)$ \citep{coughlin20}, where $Y_{\ell}^{m}(\theta,\phi)(\ell=0,1,2,...;m=-\ell,-\ell+1,...,\ell-1,\ell)$ is a spherical harmonic; the angular dependence of non-radial velocities can be written as derivatives of spherical harmonics (e.g., \citealt{cox80, hansen04}). In general, the surfaces will neither move uniformly nor oscillate in and out regularly because the displacement of a typical mass element from its unperturbed state is not radial. Thus, the motion can be complicated, for example, some regions of the surface expanding and some other region contracting at the same instant of time. In non-radial oscillations gravity is also a restoring force, implying that the spectrum will have both gravity (g-mode) and pressure (p-mode) eigenmodes (unless the envelope is modeled as a pure polytrope, in which case the vanishing of the entropy gradient removes the presence of g-modes). The detail of these considerations demand separate study and will be analyzed in future works. We also ignored the gravitational presence of the host star here, and hence the tidal force -- which will contribute an $\ell = 2$ perturbation to the envelope --  was not accounted for in our analysis. 

\section{Summary and Implications} 
\label{sec:summary}
The formation of giant planets generically involves regions where hydrogen becomes dissociated and then ionized, leading to a reduction in the average adiabatic index of the envelope. In this paper, we analyzed the stability (to radial perturbations) of hydrostatic solutions for a class of models, composed of an incompressible core and a polytropic envelope, that capture to leading-order how a reduced $\gamma$ affects planetary stability. We demonstrated that the presence of a core within a planetary interior supplies a net stabilizing effect, relative to the usual Chandresekhar criterion of $\gamma = 4/3$ that is calculated for polytropic gaseous spheres with no central mass. The stabilizing effect of the incompressible solid core is due to the fact that the core-occupied region does not contribute to the self-gravity variations, while the pressure remains approximately the same compared to the core-less configuration. Stability depends upon the core-to-total-planet mass ratio $\mu$, the core radius relative to the planetary radius $\xi_{\rm c} = r_{\rm c}/R$, and the adiabatic index $\gamma$. As shown in Figure~\ref{fig:Omega_0}, depending upon the values of these parameters the core can either fully stabilize the envelope, or lead to a strip of unstable $\gamma$ values whose upper boundary lies below the classical value of $\gamma = 4/3$.

Our model considers a single, effective adiabatic index for the entire envelope, though the motivating physics for treating small values of the adiabatic index -- partial ionization and dissociation -- implies that should, more physically, change as a function of radius and time. 
A radially varying adiabatic index implies that the background state is no longer polytropic, and buoyancy terms will modify the eigenvalues. In Appendix \ref{sec:appendix} we derive the eigenvalue equation that accounts for the effects of a time-dependent (perturbed) adiabatic index, and demonstrate that such effects can be quite important in determining the stability of a planetary envelope.  
We have also adopted inner boundary conditions for the envelope that correspond to a strictly incompressible core. In reality, rocky cores are compressible (e.g., \citealt{Rav17}) and may not have a sharp boundary \citep{Wahl17,Stevenson20}, and this finite compressibility becomes increasingly important as the majority of the gas piles up near the core surface (as occurs for the small-$\gamma$ solutions and that eventually inhibits the formation of the Chandrasekhar-like instability). When the core is compressible, its stabilizing influence on the envelope lessens, as more mass is able to penetrate deeper into the interior of the planet and increase the perturbation to self-gravity (which is ultimately responsible for generating the instability). At the outer envelope boundary, we have ignored any effects that the ambient disc gas might have on either the stability of the envelope to radial perturbations, or to the generation of Kelvin-Helmholtz-like interface instabilities alongside the Chandrasekhar-like instability analyzed here. 

The possibility of dynamical instability in the envelope of planets that form via core accretion is distinct from the existence of a maximum (or ``critical") core mass within that theory. Generically, in core accretion, solutions to the planetary structure exhibit a maximum in a plot of core mass versus total mass \citep{perri74,cameron78,hiroshi80,pollack96,Papaloizou99}. This class of solution is derived by matching the interior structure to background disc conditions at the Bondi or Hill radius, in contrast to the zero-pressure outer boundary conditions considered here. \citet{Papaloizou99} included detailed structure calculations (with energy transport by both radiation and mixing length theory convection), as well as a detailed disc model to specify how planet envelope conditions vary with disc radius. They noted that the core instability -- which  occurs in steady state calculations -- would not directly cause dynamical collapse, but would instead give an enhanced rate of cooling and Kelvin-Helmholtz contraction, as in the time-dependent structure calculations of \citet{Boden86} and as noted in the introduction. Simpler adiabatic models of the ``core accretion instability" show that it does depend on the adiabatic index \citep[e.g.][]{Bethune19}, but there is no straightforward reason to expect that the two types of instability would show similar trends or thresholds in $\gamma$.

The outcome of a Chandrasekhar-like dynamical instability, were it to occur in a forming giant planet, cannot be addressed in linear theory. It appears likely, however, that any nascent collapse would change the interior density and temperature sufficiently as to restore a stable value of $\gamma$. As a consequence, we expect the maximum degree of radial compression of the envelope to be bounded, and gas that initially collapses will rebound and exhibit a large-scale oscillation. In principle, this could lead to the formation of an outward-propagating shock wave at some depth in the interior of the planetary envelope. Depending on its strength, this shock could then eject a fraction of the planetary envelope. Alternatively, the planet could exhibit ``breathing modes,'' or large-amplitude oscillations, formed analogously to the oscillations of classical Cepheids {\citep{1991Icar...91...53W}}.

Our model does not include the accretion of material onto the core or the gaseous envelope. Accounting for the accretion of mass onto the core, and the corresponding loss of mass from the envelope, could be relatively simply accomplished by changing the inner boundary condition from $f_1(\xi_{\rm c}) = 0$ to some non-zero value. The change in the envelope mass then comes out of the continuity equation self-consistently by integrating from $\xi_{\rm c}$ to $\xi$ and imposing a non-zero $f_1(\xi_{\rm c})$, and there would then be an additional perturbation that arises from the time-dependent mass of the core (which arises from the finite flux at the core radius) that appears in the momentum equation. This change in the boundary condition would likely serve to further stabilize the envelope, as self-gravitating mass is lost to the core, the time dependence of which serves as a source term in the equations (and hence doesn't affect the eigenvalues). Alternatively, one could simply impose a time-dependent core mass and neglect the complicating issue of self-consistently accounting for the flux from the envelope. In this case the eigenvalues are unaltered as the time-dependent core mass is just a source term in the equations.

Allowing for accretion onto the planet from the surrounding disc is not as straightforward because our hydrostatic solutions terminate at a radius where the density is zero. To maintain a finite mass flux at the surface would therefore require an infinite velocity there, which, in addition to being obviously non-physical, cannot be done self-consistently in our model where the velocity is considered a perturbation. One would therefore need to change the background solution such that either the gaseuos envelope extends out indefinitely, or that the envelope terminates at a location with finite density and that is not a contact discontinuity (to permit accretion). A natural way to do the latter would be to join the envelope onto an accretion shock, meaning that the hydrostatic nature of the envelope would only be approximate\footnote{There is a time-steady, self-similar solution that does this and that has been studied in the context of neutron star accretion; see \citet{Houck92, Blondin03}.}. Doing the former could be possible if the adiabatic index is small enough (e.g., $\gamma = 1.2$ in the core-less case leads to an envelope of infinite radius), though the eigenvalues describing the perturbations to such a system would be continuous, further complicating the analysis. In either case the problem differs substantially from the one considered here, and the eigenvalues would likewise be quite different. For example, a non-zero initial velocity gradient in the unperturbed solution leads to complex eigenvalues because the operator equation is no longer Hermitian (e.g., \citealt{Coughlin19}).

Our stability maps -- or more refined versions of them that included some of the neglected effects discussed above -- would need to be combined with planetary evolution models to fully assess whether dynamical instability occurs during giant planet formation. It is immediately obvious, however, that the formation of a Jupiter-like planet via a standard core accretion channel is rendered highly stable by the presence of its core. The diversity of proto-planetary interior structures that can be formed, whether by standard core accretion, its variants, or via gravitational instability, is broad. It remains possible that the evolutionary tracks of some giant proto-planets cross into the instability strip that we have defined, leading to potentially observable time-dependent dynamics.

\section*{Data availability}
Code to reproduce the results in this paper is available upon reasonable request to the corresponding author.

\section*{Acknowledgements}
We thank Sabina Sagynbayeva for helpful discussions. 
ERC acknowledges support from the National Science Foundation through grant AST-2006684.  ANY and PJA acknowledge support from NASA TCAN award 80NSSC19K0639.  ANY acknowledges support from NASA through grant NNX17AK59G. 

%\appendix
\appendix
\section{%Time-
Varying adiabatic index}
\label{sec:appendix}

When the adiabatic index of the fluid is non-uniform and variable in time, the entropy equation takes a slightly different form that incorporates derivatives of the adiabatic index. In particular, if we start with the gas-energy equation in spherical symmetry,

\begin{equation}
    \frac{de}{dt}+\left(e+p\right)\frac{1}{r^2}\frac{\partial}{\partial r}\left[r^2 v\right] = 0,
\end{equation}
where $d/dt = \partial/\partial t+v\partial/\partial r$ is the advective derivative, then using the continuity equation turns this into

\begin{equation}
    \frac{de}{dt}-\left(e+p\right)\frac{d}{dt}\ln\rho = 0.
\end{equation}
We now define

\begin{equation}
    e = \frac{1}{\gamma-1}p,
\end{equation}
and let $\gamma$ (the adiabatic index) vary with both space and time. Then the previous equation becomes, after some algebraic rearranging,
\begin{equation}
    \frac{d}{dt}\ln\left[\frac{1}{\gamma-1}\frac{p}{\rho^{\gamma}}\right]+\ln\rho\frac{d\gamma }{dt} = 0. \label{entropy}
\end{equation}

We now introduce the non-dimensionalized density, pressure, and velocity as we did in Section \ref{sec:equations} and we perturb the variables about a background, hydrostatic state. We also allow variation in the adiabatic index and therefore write

\begin{equation}
    \gamma = \gamma_0(\xi)+\gamma_1(\xi,\tau),
\end{equation}

 Since the continuity and momentum equations are unaltered, the linearized versions of these equations are also unchanged from those found in Section \ref{sec:perturbations} (specifically Equations \ref{eq:perturbedconm} and \ref{eq:perturbedlinearmomentum}). The linearized entropy equation is recovered by inserting our definitions for the velocity, density, and pressure into Equation \eqref{entropy}, changing variables to $\tau$ and $\xi$, and maintaining first-order terms. The result is
 \begin{align}
 \begin{split}
         \frac{\partial}{\partial \tau}\left[\frac{h_1}{h_0}-\frac{\gamma_0g_1}{g_0}\right]+f\bigg(\frac{1}{h_0}\diffp{h_0}{\xi}-\frac{\gamma_0}{g_0}\diffp{g_0}{\xi}-g_0\diffp{\gamma_0}{\xi}\bigg)\\
         -\left(4-3\gamma_0\right)V =\bigg( \frac{1}{\gamma_0-1}-\ln g_0\bigg)\frac{\partial \gamma_1}{\partial \tau}+(f-V\xi)\frac{1}{\gamma_0-1}\diffp{\gamma_0}{\xi}.
\end{split}
\label{eq:nentropy}
 \end{align}
 
We notice now the entropy equation is indeed modified accommodating the variability of the adiabatic index $\gamma$. The Equation \ref{eq:nentropy} collapses into Equation \eqref{eq:perturbedentropy} when the gradient of entropy vanishes.We can now take Laplace transformation of our perturbed equations and combine them into a single, second-order equation for $\tilde{f}_1$; upon dividing by $\tilde{V}$ -- the Laplace transform of the velocity of the surface relative to the escape speed -- this equation for the eigenmodes is
\begin{multline}
    \sigma_n^2\tilde{f}_1+\frac{1}{g_0^2\xi^2}\diffp{h_0}{\xi}(\gamma_0-1)\diffp{}{\xi} \bigg((\xi-\tilde{f}_1)\diffp{m_0}{\xi}\bigg)+\\
    \frac{h_0}{g_0}\diffp{}{\xi}\bigg(\frac{\gamma_0}{g_0 \xi^2}(\xi-\tilde{f}_1)\diffp{m_0}{\xi}\bigg)=-\frac{1-\mu}{\xi^2}(\xi-\tilde{f}_1)\diffp{m_0}{\xi}\\
    -\frac{1}{g_0}\diffp{}{\xi}\Bigg[\tilde{f}_1\Bigg(\diffp{\gamma_0}{\xi}\bigg(\frac{1}{\gamma_0-1}+g_0 h_0\bigg)
    -\bigg(\diffp{h_0}{\xi}-\frac{\gamma_0 h_0}{g_0} \diffp{g_0}{\xi}\bigg)\Bigg)+\\
    h_0\bigg(\sigma_n \bigg(\frac{1}{\gamma_0-1}-\ln g_0\bigg)\frac{\tilde{\gamma}_1}{\tilde{V}}
    -\xi\frac{1}{\gamma_0-1}\diffp{\gamma_0}{\xi}+(4-3\gamma_0)\bigg)\Bigg]
    \label{eigenmodeeq}
\end{multline}

Without an additional expression that relates the perturbation to the adiabatic index to the other fluid variables, which could, for example, come from a microphysical model, we cannot make further progress on Equation \eqref{eigenmodeeq}. However, in general we expect changes in the adiabatic index to be most sensitive to changes in the gas temperature; if we therefore assume that the fractional change in the adiabatic index, $\gamma_1/\gamma_0$, is proportional to the fractional change in the temperature, $T_1/T_0$, then we can use our definitions of the fluid quantities to construct an additional relationship among $\tilde{\gamma}_1$ and the Laplace-transformed (dimensionless) fluid variables. The result is that the eigenmode equation is no longer Hermitian, and hence the eigenvalue squares $\sigma^2$ are no longer (in general) purely real. Time-dependent changes in the adiabatic index of the fluid can therefore generate distinct instabilities in the planetary interior, as is derived (much) more rigorously in \citet{cox80}. We also see that, in the area of interest here where $\gamma_0$ is close to one, the importance of time-dependent changes to the adiabatic index is amplified by the factor of $1/(\gamma_0-1)$ on the right-hand side of Equation \eqref{eigenmodeeq}. 
Changes to the adiabatic index can be quite important for understanding the generic stability of a giant-planet envelope.

\bibliographystyle{mnras}
\bibliography{planet}

\begin{thebibliography}{}
\makeatletter
\relax
\def\mn@urlcharsother{\let\do\@makeother \do\$\do\&\do\#\do\^\do\_\do\%\do\~}
\def\mn@doi{\begingroup\mn@urlcharsother \@ifnextchar [ {\mn@doi@}
  {\mn@doi@[]}}
\def\mn@doi@[#1]#2{\def\@tempa{#1}\ifx\@tempa\@empty \href
  {http://dx.doi.org/#2} {doi:#2}\else \href {http://dx.doi.org/#2} {#1}\fi
  \endgroup}
\def\mn@eprint#1#2{\mn@eprint@#1:#2::\@nil}
\def\mn@eprint@arXiv#1{\href {http://arxiv.org/abs/#1} {{\tt arXiv:#1}}}
\def\mn@eprint@dblp#1{\href {http://dblp.uni-trier.de/rec/bibtex/#1.xml}
  {dblp:#1}}
\def\mn@eprint@#1:#2:#3:#4\@nil{\def\@tempa {#1}\def\@tempb {#2}\def\@tempc
  {#3}\ifx \@tempc \@empty \let \@tempc \@tempb \let \@tempb \@tempa \fi \ifx
  \@tempb \@empty \def\@tempb {arXiv}\fi \@ifundefined
  {mn@eprint@\@tempb}{\@tempb:\@tempc}{\expandafter \expandafter \csname
  mn@eprint@\@tempb\endcsname \expandafter{\@tempc}}}

\bibitem[\protect\citeauthoryear{Adams et~al.}{Adams et~al.}{1989}]{adams89}
Adams F.~C.,  et~al., 1989, \mn@doi [Astrophysical Journal]
  {doi:10.1086/168187}, 347, 959

\bibitem[\protect\citeauthoryear{{B{\'e}thune}}{{B{\'e}thune}}{2019}]{Bethune19}
{B{\'e}thune} W.,  2019, \mn@doi [\mnras] {10.1093/mnras/stz2796}, \href
  {https://ui.adsabs.harvard.edu/abs/2019MNRAS.490.3144B} {490, 3144}

\bibitem[\protect\citeauthoryear{{Blondin}, {Mezzacappa}  \&
  {DeMarino}}{{Blondin} et~al.}{2003}]{Blondin03}
{Blondin} J.~M.,  {Mezzacappa} A.,   {DeMarino} C.,  2003, \mn@doi [\apj]
  {10.1086/345812}, \href
  {https://ui.adsabs.harvard.edu/abs/2003ApJ...584..971B} {584, 971}

\bibitem[\protect\citeauthoryear{Bodenheimer \& Pollack}{Bodenheimer \&
  Pollack}{1986}]{Boden86}
Bodenheimer P.,  Pollack J.~B.,  1986, \mn@doi []
  {https://doi.org/10.1016/0019-1035(86)90122-3}, 67, 391

\bibitem[\protect\citeauthoryear{Boss}{Boss}{1997}]{Boss97}
Boss A.~P.,  1997, \mn@doi [Science] {doi:10.1126/science.276.5320.1836}, 276,
  1836

\bibitem[\protect\citeauthoryear{{Cameron}}{{Cameron}}{1978}]{cameron78}
{Cameron} A.~G.~W.,  1978, \mn@doi [Moon and Planets] {10.1007/BF00896696},
  \href {https://ui.adsabs.harvard.edu/abs/1978M&P....18....5C} {18, 5}

\bibitem[\protect\citeauthoryear{Chandrasekhar \& Milne}{Chandrasekhar \&
  Milne}{1933}]{chandra33}
Chandrasekhar S.,  Milne E.~A.,  1933, \mn@doi [Monthly Notices of the Royal
  Astronomical Society, Volume 93, Issue 5, March 1933, Pages 390–406,]
  {https://doi.org/10.1093/mnras/93.5.390}, 93

\bibitem[\protect\citeauthoryear{Chavanis}{Chavanis}{2002}]{Chavanis}
Chavanis P.~H.,  2002, \mn@doi [Astronomy and Astrophysics]
  {10.1051/0004-6361:20020306}, 386, 732

\bibitem[\protect\citeauthoryear{Chiang \& Youdin}{Chiang \&
  Youdin}{2010}]{doi:10.1146/annurev-earth-040809-152513}
Chiang E.,  Youdin A.,  2010, \mn@doi [Annual Review of Earth and Planetary
  Sciences] {10.1146/annurev-earth-040809-152513}, 38, 493

\bibitem[\protect\citeauthoryear{{Coughlin} \& {Nixon}}{{Coughlin} \&
  {Nixon}}{2020}]{coughlin20}
{Coughlin} E.~R.,  {Nixon} C.~J.,  2020, \mn@doi [\apjs]
  {10.3847/1538-4365/ab77c2}, \href
  {https://ui.adsabs.harvard.edu/abs/2020ApJS..247...51C} {247, 51}

\bibitem[\protect\citeauthoryear{{Coughlin}, {Ro}  \& {Quataert}}{{Coughlin}
  et~al.}{2019}]{Coughlin19}
{Coughlin} E.~R.,  {Ro} S.,   {Quataert} E.,  2019, \mn@doi [\apj]
  {10.3847/1538-4357/ab09ec}, \href
  {https://ui.adsabs.harvard.edu/abs/2019ApJ...874...58C} {874, 58}

\bibitem[\protect\citeauthoryear{{Cox}}{{Cox}}{1980}]{cox80}
{Cox} J.~P.,  1980, {Theory of stellar pulsation}

\bibitem[\protect\citeauthoryear{Ginzburg \& Chiang}{Ginzburg \&
  Chiang}{2019}]{ginz19}
Ginzburg S.,  Chiang E.,  2019, \mn@doi [MNRAS] {10.1093/mnras/stz2901}, 490,
  4334

\bibitem[\protect\citeauthoryear{Goldreich, Lithwick  \& Sari}{Goldreich
  et~al.}{2004}]{Goldreich_2004}
Goldreich P.,  Lithwick Y.,   Sari R.,  2004, \mn@doi [The Astrophysical
  Journal] {10.1086/423612}, 614, 497

\bibitem[\protect\citeauthoryear{{Hansen}, {Kawaler}  \& {Trimble}}{{Hansen}
  et~al.}{2004}]{hansen04}
{Hansen} C.~J.,  {Kawaler} S.~D.,   {Trimble} V.,  2004, {Stellar interiors :
  physical principles, structure, and evolution}

\bibitem[\protect\citeauthoryear{Harris}{Harris}{1978}]{harris78}
Harris A.~W.,  1978, Lunar and Planetary Science Conference, p.~459

\bibitem[\protect\citeauthoryear{{Heinisch} et~al.,}{{Heinisch}
  et~al.}{2019}]{2019A&A...630A...2H}
{Heinisch} P.,  et~al., 2019, \mn@doi [\aap] {10.1051/0004-6361/201833889},
  \href {https://ui.adsabs.harvard.edu/abs/2019A&A...630A...2H} {630, A2}

\bibitem[\protect\citeauthoryear{Helled \& Stevenson}{Helled \&
  Stevenson}{2017}]{Rav17}
Helled R.,  Stevenson D.,  2017, \mn@doi [The Astrophysical Journal Letters]
  {10.3847/2041-8213/aa6d08}, \href
  {https://iopscience.iop.org/article/10.3847/2041-8213/aa6d08} {840}

\bibitem[\protect\citeauthoryear{{Houck} \& {Chevalier}}{{Houck} \&
  {Chevalier}}{1992}]{Houck92}
{Houck} J.~C.,  {Chevalier} R.~A.,  1992, \mn@doi [\apj] {10.1086/171679},
  \href {https://ui.adsabs.harvard.edu/abs/1992ApJ...395..592H} {395, 592}

\bibitem[\protect\citeauthoryear{{Ikoma}, {Nakazawa}  \& {Emori}}{{Ikoma}
  et~al.}{2000}]{2000ApJ...537.1013I}
{Ikoma} M.,  {Nakazawa} K.,   {Emori} H.,  2000, \mn@doi [\apj]
  {10.1086/309050}, \href
  {https://ui.adsabs.harvard.edu/abs/2000ApJ...537.1013I} {537, 1013}

\bibitem[\protect\citeauthoryear{{Johansen}, {Blum}, {Tanaka}, {Ormel},
  {Bizzarro}  \& {Rickman}}{{Johansen} et~al.}{2014}]{2014prpl.conf..547J}
{Johansen} A.,  {Blum} J.,  {Tanaka} H.,  {Ormel} C.,  {Bizzarro} M.,
  {Rickman} H.,  2014, in {Beuther} H.,  {Klessen} R.~S.,  {Dullemond} C.~P.,
  {Henning} T.,  eds, Protostars and Planets VI. p.~547 (\mn@eprint {arXiv}
  {1402.1344}), \mn@doi{10.2458/azu\_uapress\_9780816531240-ch024}

\bibitem[\protect\citeauthoryear{{Kratter}}{{Kratter}}{2011}]{2011ASPC..447...47K}
{Kratter} K.~M.,  2011, in {Schmidtobreick} L.,  {Schreiber} M.~R.,   {Tappert}
  C.,  eds,  Astronomical Society of the Pacific Conference Series Vol. 447,
  Evolution of Compact Binaries. p.~47 (\mn@eprint {arXiv} {1109.3740})

\bibitem[\protect\citeauthoryear{{Lambrechts} \& {Lega}}{{Lambrechts} \&
  {Lega}}{2017}]{2017A&A...606A.146L}
{Lambrechts} M.,  {Lega} E.,  2017, \mn@doi [\aap]
  {10.1051/0004-6361/201731014}, \href
  {https://ui.adsabs.harvard.edu/abs/2017A&A...606A.146L} {606, A146}

\bibitem[\protect\citeauthoryear{{Larson}}{{Larson}}{1969}]{larson69}
{Larson} R.~B.,  1969, \mn@doi [\mnras] {10.1093/mnras/145.3.271}, \href
  {https://ui.adsabs.harvard.edu/abs/1969MNRAS.145..271L} {145, 271}

\bibitem[\protect\citeauthoryear{Lee \& Chiang}{Lee \& Chiang}{2015}]{Lee_2015}
Lee E.~J.,  Chiang E.,  2015, \mn@doi [The Astrophysical Journal]
  {10.1088/0004-637x/811/1/41}, 811, 41

\bibitem[\protect\citeauthoryear{{Lissauer}, {Hubickyj}, {D'Angelo}  \&
  {Bodenheimer}}{{Lissauer} et~al.}{2009}]{2009Icar..199..338L}
{Lissauer} J.~J.,  {Hubickyj} O.,  {D'Angelo} G.,   {Bodenheimer} P.,  2009,
  \mn@doi [\icarus] {10.1016/j.icarus.2008.10.004}, \href
  {https://ui.adsabs.harvard.edu/abs/2009Icar..199..338L} {199, 338}

\bibitem[\protect\citeauthoryear{Liu}{Liu}{1996}]{Liu96}
Liu F.~K.,  1996, \mn@doi [Monthly Notices of the Royal Astronomical Society]
  {https://doi.org/10.1093/mnras/281.4.1197}, 281

\bibitem[\protect\citeauthoryear{{Liu}, {Ormel}  \& {Johansen}}{{Liu}
  et~al.}{2019}]{2019A&A...624A.114L}
{Liu} B.,  {Ormel} C.~W.,   {Johansen} A.,  2019, \mn@doi [\aap]
  {10.1051/0004-6361/201834174}, \href
  {https://ui.adsabs.harvard.edu/abs/2019A&A...624A.114L} {624, A114}

\bibitem[\protect\citeauthoryear{Lubow, Seibert  \& Artymowicz}{Lubow
  et~al.}{1999}]{Lubow_1999}
Lubow S.~H.,  Seibert M.,   Artymowicz P.,  1999, \mn@doi [The Astrophysical
  Journal] {10.1086/308045}, 526, 1001–1012

\bibitem[\protect\citeauthoryear{Machida, Kokubo, ichiro Inutsuka  \&
  Matsumoto}{Machida et~al.}{2008}]{Machida_2008}
Machida M.~N.,  Kokubo E.,  ichiro Inutsuka S.,   Matsumoto T.,  2008, \mn@doi
  [The Astrophysical Journal] {10.1086/590421}, 685, 1220

\bibitem[\protect\citeauthoryear{Mizuno}{Mizuno}{1980}]{hiroshi80}
Mizuno H.,  1980, \mn@doi [Progress of Theoretical Physics, Volume 64, Issue 2,
  August 1980, Pages 544–557] {https://doi.org/10.1143/PTP.64.544}, 64

\bibitem[\protect\citeauthoryear{{Nixon}, {King}  \& {Pringle}}{{Nixon}
  et~al.}{2018}]{nixon18}
{Nixon} C.~J.,  {King} A.~R.,   {Pringle} J.~E.,  2018, \mn@doi [\mnras]
  {10.1093/mnras/sty593}, \href
  {https://ui.adsabs.harvard.edu/abs/2018MNRAS.477.3273N} {477, 3273}

\bibitem[\protect\citeauthoryear{{Ormel} \& {Cuzzi}}{{Ormel} \&
  {Cuzzi}}{2007}]{2007A&A...466..413O}
{Ormel} C.~W.,  {Cuzzi} J.~N.,  2007, \mn@doi [\aap]
  {10.1051/0004-6361:20066899}, \href
  {https://ui.adsabs.harvard.edu/abs/2007A&A...466..413O} {466, 413}

\bibitem[\protect\citeauthoryear{{Papaloizou} \& {Terquem}}{{Papaloizou} \&
  {Terquem}}{1999}]{Papaloizou99}
{Papaloizou} J. C.~B.,  {Terquem} C.,  1999, \mn@doi [\apj] {10.1086/307581},
  \href {https://ui.adsabs.harvard.edu/abs/1999ApJ...521..823P} {521, 823}

\bibitem[\protect\citeauthoryear{Perri \& Cameron}{Perri \&
  Cameron}{1974}]{perri74}
Perri F.,  Cameron A. G.~W.,  1974, Icarus, 22, 416

\bibitem[\protect\citeauthoryear{{Piso} \& {Youdin}}{{Piso} \&
  {Youdin}}{2014}]{2014ApJ...786...21P}
{Piso} A.-M.~A.,  {Youdin} A.~N.,  2014, \mn@doi [\apj]
  {10.1088/0004-637X/786/1/21}, \href
  {https://ui.adsabs.harvard.edu/abs/2014ApJ...786...21P} {786, 21}

\bibitem[\protect\citeauthoryear{{Piso}, {Youdin}  \& {Murray-Clay}}{{Piso}
  et~al.}{2015}]{2015ApJ...800...82P}
{Piso} A.-M.~A.,  {Youdin} A.~N.,   {Murray-Clay} R.~A.,  2015, \mn@doi [\apj]
  {10.1088/0004-637X/800/2/82}, \href
  {https://ui.adsabs.harvard.edu/abs/2015ApJ...800...82P} {800, 82}

\bibitem[\protect\citeauthoryear{Pollack et~al.}{Pollack
  et~al.}{1996}]{pollack96}
Pollack J.~B.,  et~al., 1996, \mn@doi [] {doi:10.1006/icar.1996.0190}, 124, 62

\bibitem[\protect\citeauthoryear{Safronov}{Safronov}{1972}]{safronov72}
Safronov V.~S.,  1972, Evolution of the protoplanetary cloud and formation of
  the earth and the planets. Jerusalem, Israel Program for Scientific
  Translations.

\bibitem[\protect\citeauthoryear{Saumon, Chabrier  \& van Horn}{Saumon
  et~al.}{1995}]{saumon95}
Saumon D.,  Chabrier G.,   van Horn H.~M.,  1995, \mn@doi [Astrophysical
  Journal Supplement] {doi:10.1086/192204}, 99, 713

\bibitem[\protect\citeauthoryear{Stevenson}{Stevenson}{1982}]{STEVENSON1982755}
Stevenson D.,  1982, \mn@doi [Planetary and Space Science]
  {https://doi.org/10.1016/0032-0633(82)90108-8}, 30, 755

\bibitem[\protect\citeauthoryear{{Stevenson}}{{Stevenson}}{2020}]{Stevenson20}
{Stevenson} D.~J.,  2020, \mn@doi [Annual Review of Earth and Planetary
  Sciences] {10.1146/annurev-earth-081619-052855}, \href
  {https://ui.adsabs.harvard.edu/abs/2020AREPS..48..465S} {48, 465}

\bibitem[\protect\citeauthoryear{Tanigawa, Ohtsuki  \& Machida}{Tanigawa
  et~al.}{2012}]{Tanigawa_2012}
Tanigawa T.,  Ohtsuki K.,   Machida M.~N.,  2012, \mn@doi [The Astrophysical
  Journal] {10.1088/0004-637x/747/1/47}, 747, 47

\bibitem[\protect\citeauthoryear{{Wahl} et~al.,}{{Wahl} et~al.}{2017}]{Wahl17}
{Wahl} S.~M.,  et~al., 2017, \mn@doi [\grl] {10.1002/2017GL073160}, \href
  {https://ui.adsabs.harvard.edu/abs/2017GeoRL..44.4649W} {44, 4649}

\bibitem[\protect\citeauthoryear{Wetherill \& Stewart}{Wetherill \&
  Stewart}{1993}]{wetherill93}
Wetherill G.~W.,  Stewart G.~R.,  1993, \mn@doi [Icarus,Volume 106, Issue 1,
  November 1993, Pages 190-209] {https://doi.org/10.1006/icar.1993.1166}, 106

\bibitem[\protect\citeauthoryear{{Wuchterl}}{{Wuchterl}}{1991}]{1991Icar...91...53W}
{Wuchterl} G.,  1991, \mn@doi [\icarus] {10.1016/0019-1035(91)90125-D}, \href
  {https://ui.adsabs.harvard.edu/abs/1991Icar...91...53W} {91, 53}

\bibitem[\protect\citeauthoryear{Youdin \& Goodman}{Youdin \&
  Goodman}{2005}]{Youdin_2005}
Youdin A.~N.,  Goodman J.,  2005, \mn@doi [The Astrophysical Journal]
  {10.1086/426895}, 620, 459

\bibitem[\protect\citeauthoryear{{Youdin} \& {Kenyon}}{{Youdin} \&
  {Kenyon}}{2013}]{2013pss3.book....1Y}
{Youdin} A.~N.,  {Kenyon} S.~J.,  2013, {From Disks to Planets}.
p.~1, \mn@doi{10.1007/978-94-007-5606-9\_1}

\makeatother
\end{thebibliography}

\bsp	% typesetting comment
 \label{lastpage}\end{document}